\documentclass[12pt,preprint]{aastex}

\slugcomment{Submitted: ApJ, Sept. 2008; revised Nov. 25, 2008}

\shorttitle{Distance to NGC~0247}
\shortauthors{Madore et al }

\begin{document}
  \title{\bf The Cepheid Distance to NGC 0247}
\author{\bf Barry F.~Madore \& Wendy L. Freedman} 
\affil{The Observatories\\
Carnegie Institution of Washington\\
813 Santa Barbara Street\\
Pasadena, CA 91101\\
email: barry@ociw.edu, wendy@ociw.edu}

\author{\bf Joseph Catanzarite}
\affil{Jet Propulsion Laboratory, MS 301-486\\
480 Oak Grove Dr. \\
Pasadena, CA 91109-8099\\
email: joseph.h.catanzarite@jpl.nasa.gov}

\author{\bf Mauricio Navarrete}
\affil{Las Campanas Observatories\\
Carnegie Institution of Washington\\
La Serena, Chile\\
email: mnavarrete@lco.cl}

\begin{abstract}

We report VRI CCD observations of nine Cepheids in the South Polar
(Sculptor) Group spiral galaxy NGC~0247.  Periods of these Cepheids
range from 20 to 70 days. Over the past 20 years the very brightest
Cepheid in our sample, NGC~0247:[MF09]~C1,  has decreased its period by
6\%, faded by 0.8~mag in the V band, and become bluer by 0.23~mag in
(V-I). A multi-wavelength analysis of the Cepheid data yields a true
distance modulus of $\mu_o$ = 27.81$\pm$0.10~mag (3.36$\pm$0.16~Mpc)
with a total line-of-sight reddening of E(V-I) = 0.07$\pm$0.04~mag,
after adopting an LMC true distance modulus of 18.5~mag and reddening
of E(B-V) = 0.10~mag. These results are in excellent agreement with
other very recently published (Cepheid and TRGB) distances to
NGC~0247. Combining both Cepheid datasets gives $\mu_o =
27.85\pm0.09$~mag (3.72$\pm$0.15~Mpc) with E(V-I) = 0.11$\pm$0.03~mag.

\end{abstract}
\keywords{ Cepheids: galaxies: distances and redshifts --- galaxies:
individual (NGC~0247) --- galaxies: Sculptor Group }

\section{Introduction}   
    
In 1983 we began a ground-based project to provide a more secure
calibration of the zero-point for secondary distance indicators (such
as the Tully-Fisher relation) by building up a database of accurate
Cepheid distances to nearby spiral galaxies.  In due course the Hubble
Space Telescope was launched and other activities took precedence over
the ground-based effort. Preliminary mention of work on Cepheids
discovered in NGC~0247 was given in Freedman et al. (1988) and again
in Catanzarite, Freedman, Horowitz \& Madore (1994) but details were
never published, until now.  Here we present the photometric data and
provide a brief analysis leading to a distance determination to
NGC~0247 based on nine Cepheids discovered in NGC~0247 some
twenty-five years ago. We then go on to compare it with new
observations published by Garcia-Varela et al. (2008, hereafter
[GV08]).

\section{Observations}
Observations for this program were carried out over a span of eight
years (giving a time baseline of almost 3,000 days). Oberving began
first at the Cerro Tololo 4m, and was completed using the 2.5m duPont
telescope at Las Campanas, Chile.  Three fields of NGC~0247 were
surveyed in $B$, $V$, $R$, and $I$ filters at the Cerro~ Tololo
4m~telescope during November~1984 to November~1988.  Figure~1 shows a
photograph of NGC~0247 with the three fields delineated. The
coordinates of the CCD field centers are given in Table 1.  From the
second through the fourth year of the program data were obtained in
the service-observing mode offered at CTIO; those data were 
taken by M. Navarrete.  For most of the runs the $512\times320$
RCA chip \#5 (having a scale of 0.60~arcsec/pxl and a total
field of view of ~ 3$'$ by 5$'$ at the prime focus) was used.  In 1988
a different RCA chip (\#4) with similar characteristics was
substituted.  Exposure times for these frames were typically 400~sec
in $B$ and 300~sec in $V$, $R$ and $I$.  The frames were
bias-subtracted, flat-fielded, and defringed using standard
data-reduction packages available at Cerro Tololo.  Beginning in 1990,
the observing program for the Sculptor galaxies shifted to the duPont
2.5m telescope at the Las~Campanas~Observatory.  $BVRI$ CCD
observations covering the same three selected fields were obtained in
December 1990, in September 1991, and in October, November, and
December 1992.  Exposure times at this telescope were generally
900~sec in $B$ and 600~sec in $V$, $R$ and $I$\@.  Most of these
observations were electronically binned ($2 \times 2$) at the
telescope.  For the 1990 and 1991 runs, the FORD1 CCD chip was used.
For the October and November 1992 runs a Tektronix~CCD~chip~(TEK4) was
used; for the December 1992 runs, Tektronix CCD chips were also used
(TEK3 and TEK4)\@.  These chips each had dimensions of $2048 \times
2048$ pixels; the image scales obtained were: FORD1: 0.16~arcsec/pxl;
TEK3: 0.23~arcsec/pxl; and TEK4: 0.26~arcsec/pxl.  The survey totalled
about 250 exposures on 29 different nights over a span of eight years.
Table 2 gives a journal of the observations.


\section{Photometry Reduction and Calibration}
Photometric calibration of the CTIO frames was accomplished using E-region
standards in E1, E2, E3, E7, E8, and E9 (Graham 1984) and in SA 98
\cite{lan83}.  $BVRI$ standards were taken on 20 independent
photometric nights.  As described in Freedman et al. (1992), a check
on the external accuracy of the photometric calibration for these runs
was made by individually calibrating the frames for NGC~0300.  The
magnitudes for the brightest stars were in agreement to within
0.01-0.03$\pm$0.03 mag of the average for all filter/field
combinations. 

The CCD frames from CTIO were reduced using both DoPHOT (Schechter,
Mateo \& Saha 1993) and DAOPHOT (Stetson 1987), and cross-checked.
The unresolved background level in NGC~0247 is highly non-uniform, and
is characterized both by regions in which there are strong spiral arms
as well as relatively blank, interarm regions. In order to maximize
the detection limits of the algorithm FIND in DAOPHOT, the frames were
first median-smoothed using a 7$\times$ 7 pixel boxcar averaging
scheme, and subtracted from the original frames.

Details of the calibration process are discussed in~\cite{fre92}.  The
LCO data were reduced using a variant of the DoPHOT package
\cite{mat89}.  This version of DoPHOT uses median smoothing to
construct an initial model of the background sky before searching for
objects.  The sky model is refined after objects at the
next-to-the-lowest threshold have been found and subtracted.  The
refined sky model is then adopted as the baseline, and objects are
again found down to the lowest threshold and their PSF parameters
re-measured.  LCO frames were brought onto the CTIO calibrated
magnitude system by the following process: The $(B-V)$ color
term for the CCD chips used at LCO relative to RCA chip used in the
CTIO observations was measured and a correction was applied to the LCO
$B$ photometry (the LCO $V$ photometry had no significant $(B-V)$
color term.)  Next, the magnitude zero-point of each LCO frame was
offset to the instrumental zero-point of a fiducial CTIO frame, for
the corresponding field and filter.  The calibration transformation
derived for the fiducial CTIO frame was then applied to the LCO data.

\section{The Cepheids}
For Fields 1 and 3 all of the observations were tied to the
photometric zero point for October 7, 1988.  Observations on this
night were taken under excellent seeing conditions and had the best
photometric calibration available to us.  For Field 2, October 13,
1988 was used to calibrate the data.  To put all of the stars in each
field on the same coordinate system, all frames from each field were
spatially registered to the October 7, 1988 $V$ frame for that field,
(since that was the best or close to best $V$ night for all three
fields).  Coordinate transformations produced matches with $rms$
scatter of $\pm$0.30 pixels or better.  Calibrated, matched photometry
files containing the entire set of observations for each
field/bandpass combination were produced.  Stars with high internal
$V$ magnitude dispersion were then identified as described in
\cite{fre94}.  These variable candidates were then subjected to a
further test: a star was flagged as a Cepheid candidate only if the
histogram of its magnitudes was consistent with a uniform magnitude
histogram, as expected for Cepheids.  All three fields were searched
for variables down to a signal-to-noise level of 1.5$\sigma$.  The $V$
photometric data for each candidate was then phased to the twelve
periods (in the range of 1 to 100 days) with the lowest phase
dispersions, using a routine based on the Lafler-Kinman algorithm
(\cite{laf65}).  The $V$ light curves were then visually inspected and
the best period was selected.  Calibrated $B$, $R$, and $I$
observations were then phased to this adopted period and the
multi-wavelength light curves were inspected for consistency.
Candidates with strong correlation of phase and amplitude between
their $BVRI$ light curves, having well-determined periods, mean colors
and well-sampled light curves characteristic of known Cepheids were
then identified as Cepheids.  Each of these stars was then visually
inspected in the best image frame to check for nearby companions.
Nine Cepheids in total made it through the selection procedure.  All
were found in $BVRI$, with the exception of NGC~0247:[MF09]~C9, which
was was too faint to be recovered on the $I$ frames.  The positions
for the nine Cepheids in our sample are given in Table~3, the first 6
of which are mapped over from [GV08], with the positions for C7, C8
and C9 being on that system but having lower precision.  The
individual Cepheid observations are presented in Tables~4 through
12. The light curves are shown in Figure 2.  The time-averaged
properties of the individual Cepheids are listed in Table~4.

\begin{figure}
\resizebox{0.32\hsize}{!}{\includegraphics[width=0.4\textwidth, angle=-90]{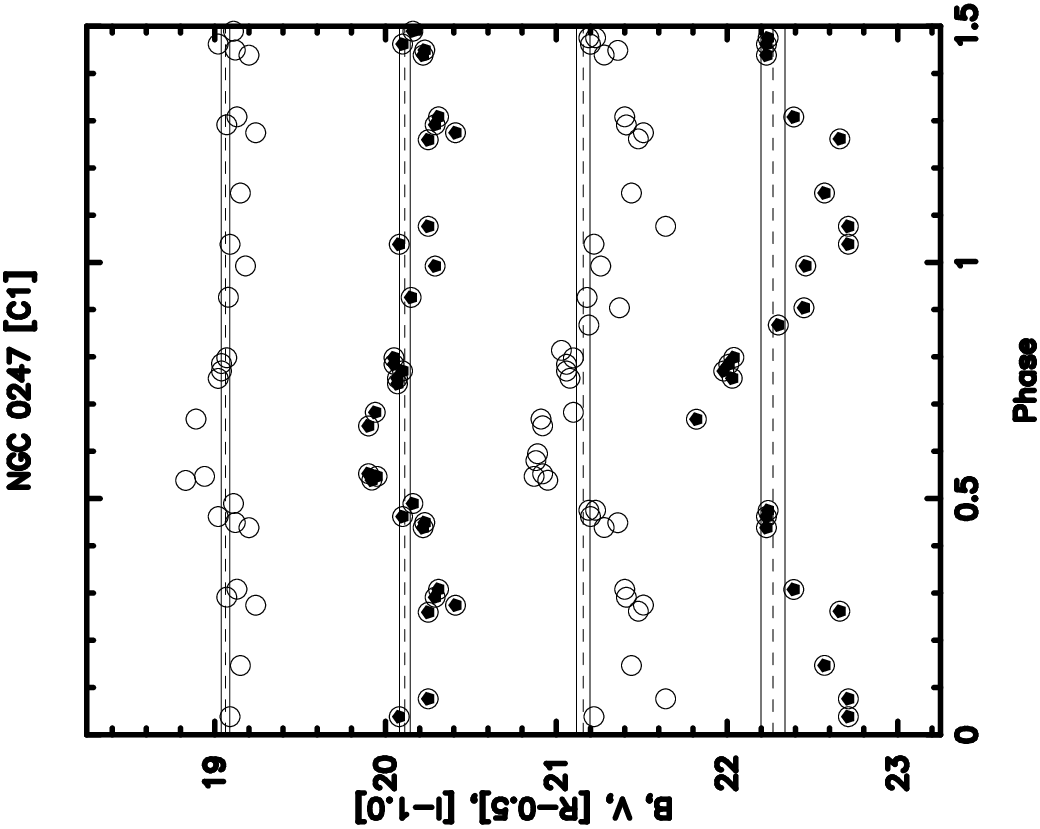}}\hspace{0.02cm}
\resizebox{0.32\hsize}{!}{\includegraphics[angle=-90]{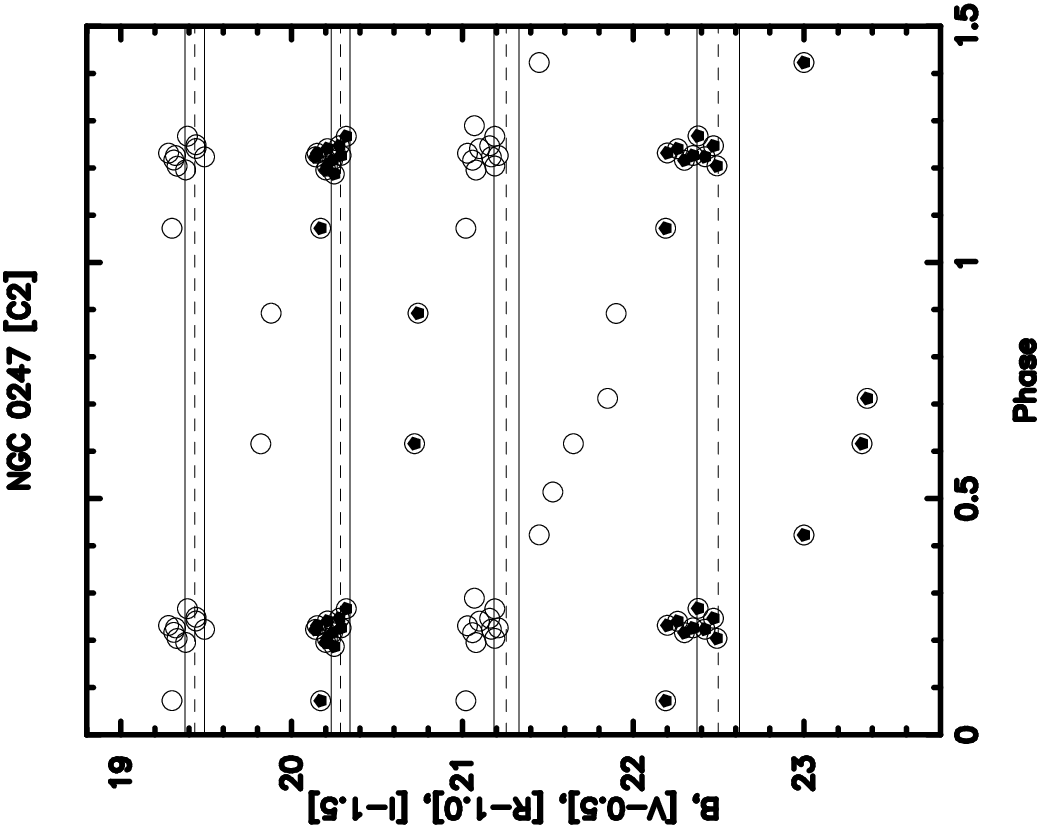}}
\resizebox{0.32\hsize}{!}{\includegraphics[angle=-90]{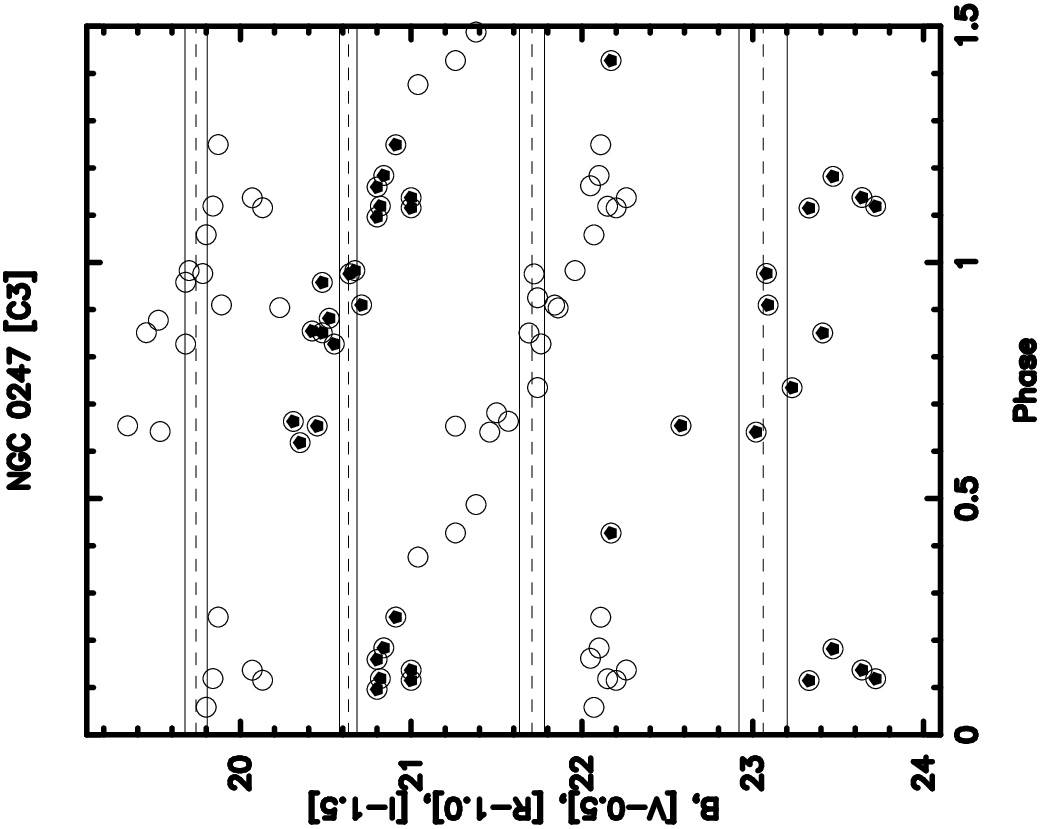}}\\

\resizebox{0.32\hsize}{!}{\includegraphics[width=0.4\textwidth, angle=-90]{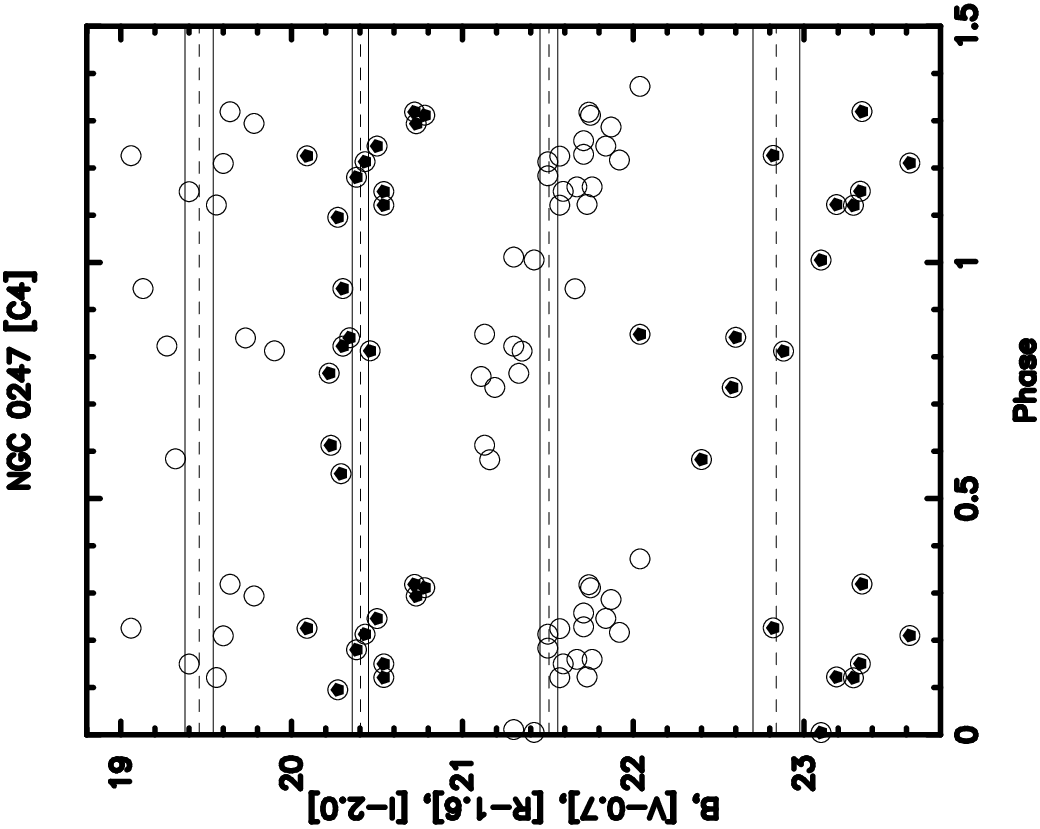}}\hspace{0.02cm}
\resizebox{0.32\hsize}{!}{\includegraphics[angle=-90]{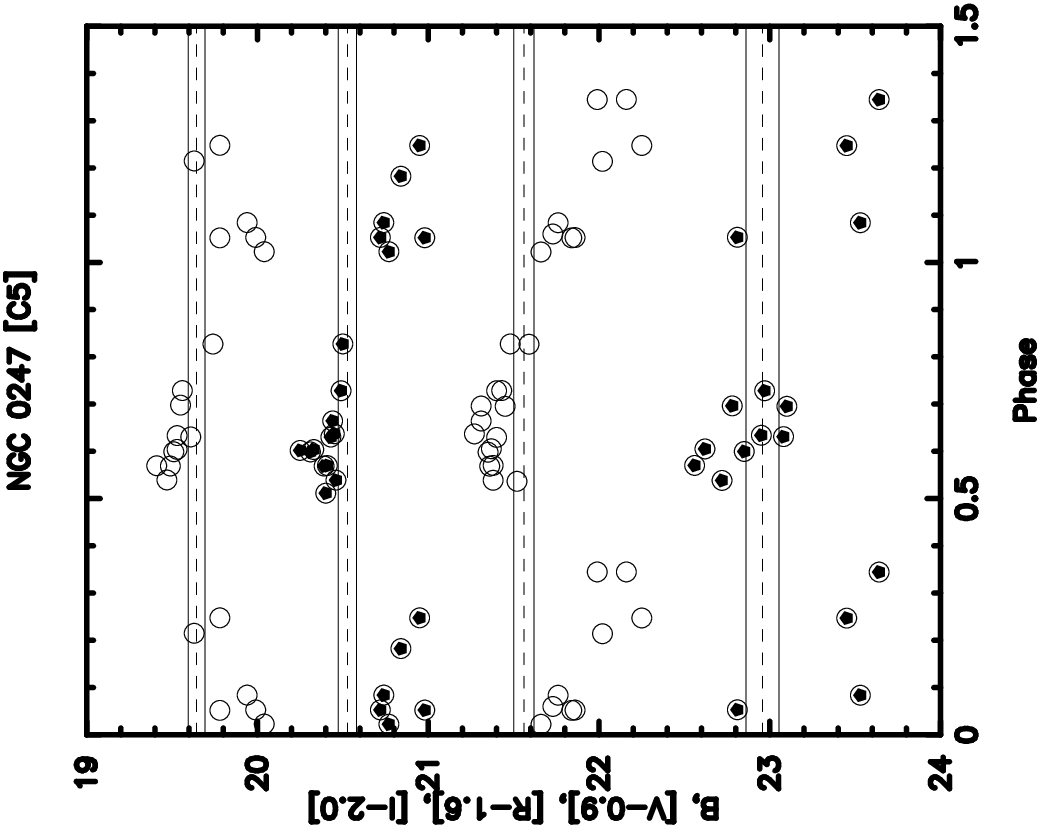}}
\resizebox{0.32\hsize}{!}{\includegraphics[angle=-90]{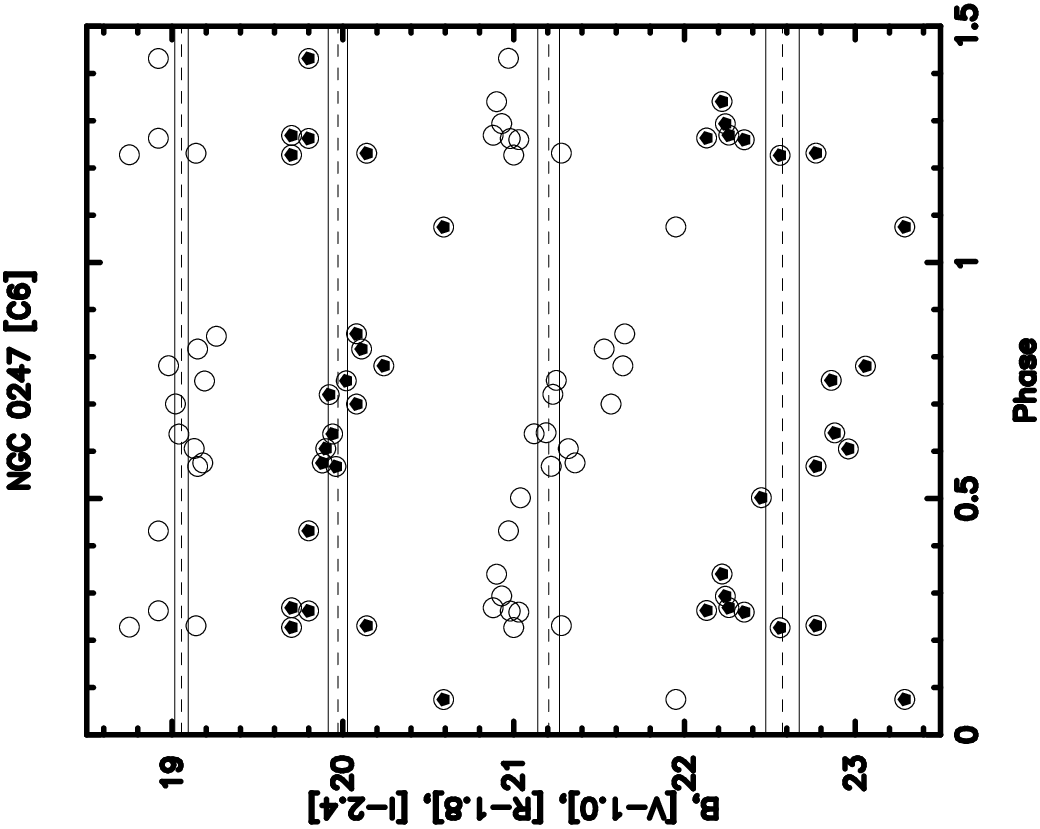}}\\

\resizebox{0.32\hsize}{!}{\includegraphics[width=0.4\textwidth, angle=-90]{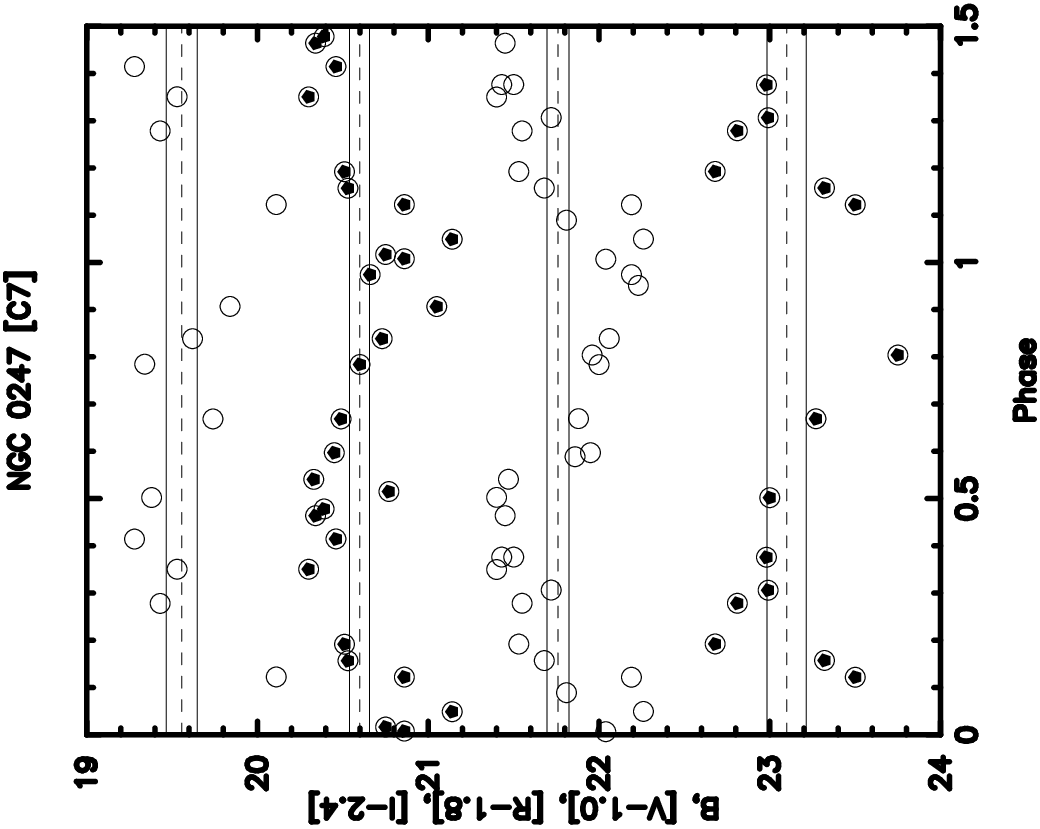}}\hspace{0.02cm}
\resizebox{0.32\hsize}{!}{\includegraphics[angle=-90]{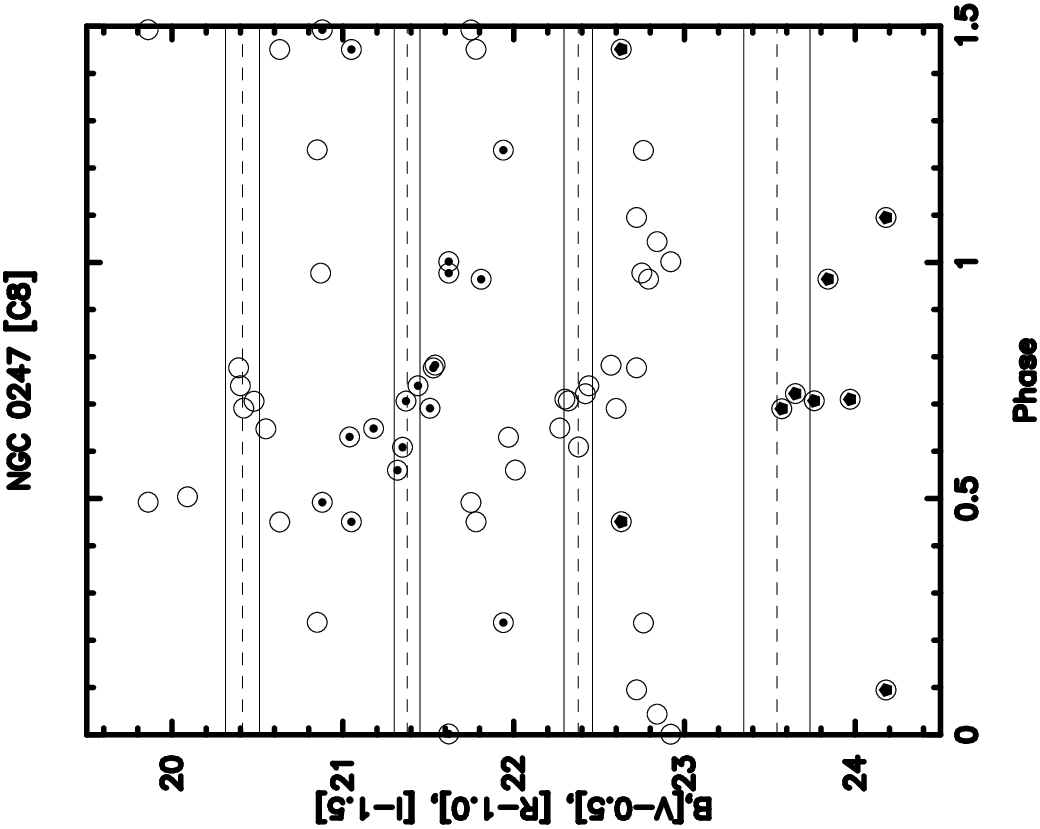}}
\resizebox{0.32\hsize}{!}{\includegraphics[angle=-90]{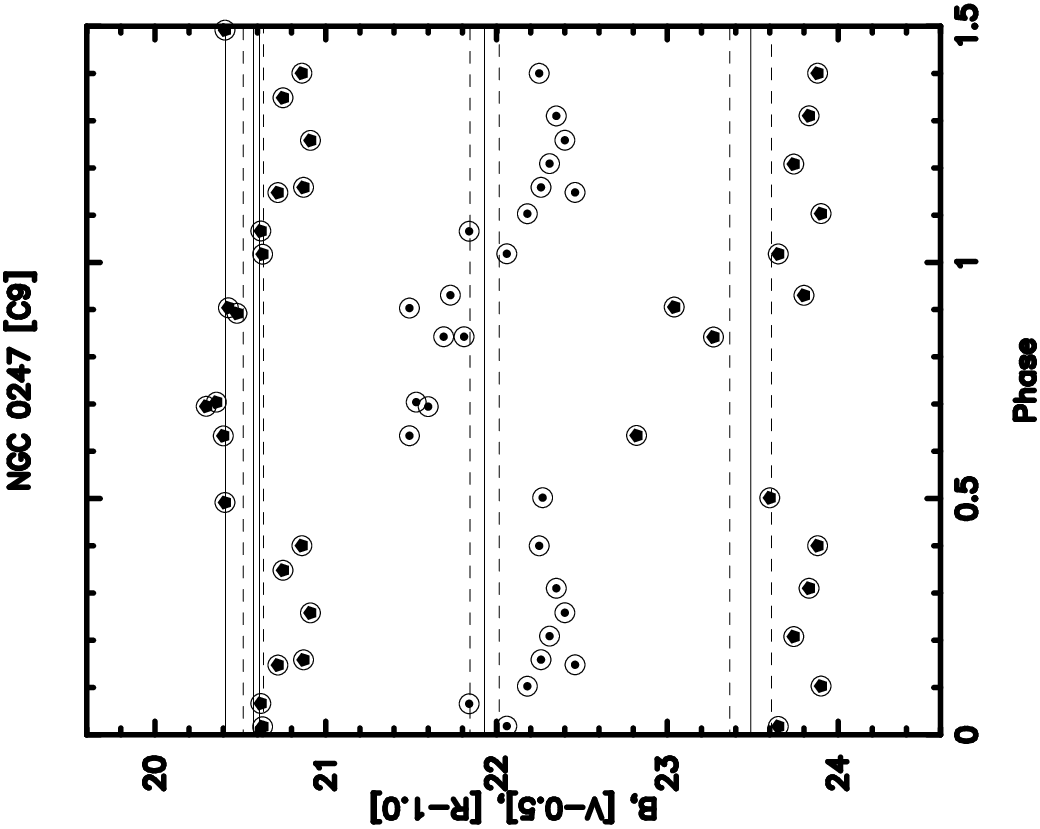}}\\
\caption{$BVRI$ lightcurves for the individual Cepheids.  The plotted
magnitude range is 5~mag in all cases. Magnitude offsets, applied, to
make the lightcurves individually more visible, are given in the
vertical-axis labels. In order from top to bottom the lightcurves are
$I$, $R$, $V$ and $B$.}
\end{figure}

\subsection{Other Variable Stars Found in NGC~0247}
Five variable stars which could not be classified as Cepheids were
also discovered.  These stars are well-isolated, their photometry is
well-measured by DoPHOT, and they have extremely strong $BVRI$
correlation.  Three of them are very red, and have light curves with a
``square wave'' shape.  If they are eclipsing variables then we have
probably been unable to determine the periods correctly.  A fourth
object has a light curve with the right shape to be a Cepheid, but is
extremely red.  The fifth may be a Cepheid with an uncharacteristic
light curve.  The properties of these stars are summarized in Table
13.

\section{The Distance to NGC~0247}

A comprehensive review of previously published distance estimates to
NGC~0247 is given in [GV08].\footnote{In addition, an updated, on-line
compilation of distances to nearby galaxies, including NGC~0247, is
available through the {\it NASA/IPAC Extragalactic Database} at the
following URL: http://nedwww.ipac.caltech.edu/level5/NED1D/intro.html.}
In that paper the authors also present their new $VI$ observations of
23 Cepheids in the period range 17 to 131 days.  Based on those two
colors they derive a true distance modulus of 27.80$\pm$0.09~mag
(3.6~Mpc) tied to an LMC true distance modulus of 18.50~mag, as also
adopted in this paper. Another important independent distance
measurement to NGC~0247 worth noting here, because of its comparably
high precision, is the tip of the red giant branch (TRGB) distance
modulus ($\mu_o$ = 27.81~mag or 3.65~Mpc) published by Karachentsev et
al. (2006).

\subsection{Discussion of Data}

The detected Cepheids at $B$ lie closer to the photometry limits than
at $V$, $R$, or $I$; furthermore deriving a stable zero-point for that
bandpass was found to be problematic.  As such, the $B$ data were used
to confirm the periods adopted here, but because of signal-to-noise
and other calibration problems we do not use the $B$-band data further
in this paper. The $B$-band data are listed in this paper, but readers
are strongly warned against using it for anything quantitative until a
proper calibration is found.  The time-averaged data for our 9
Cepheids are given in Table 14.  The periods cited there were derived
from these data alone, but will be updated later in the paper when we
consider  a merger with the [GV08] sample.

As can be seen in Figure 12, the NGC~0247 PL relations in $V$, $R$,
and $I$, have smaller observed dispersions than the fiducial LMC PL
relations whose 2-sigma widths are shown by the dashed lines.  The
small observed dispersion is presumably due to small number
statistics, but it could also be signalling a slight bias in the
sample.  If the instability strip is not being fully sampled we cannot
be sure that these Cepheids properly reflect the mean. An external
check with the results of [GV08] (Section 6 below) would suggest that
that bias (between samples) is at or below the 0.1 mag level.

\begin{figure}
\begin{center}
\includegraphics[angle=-90, scale=0.65]{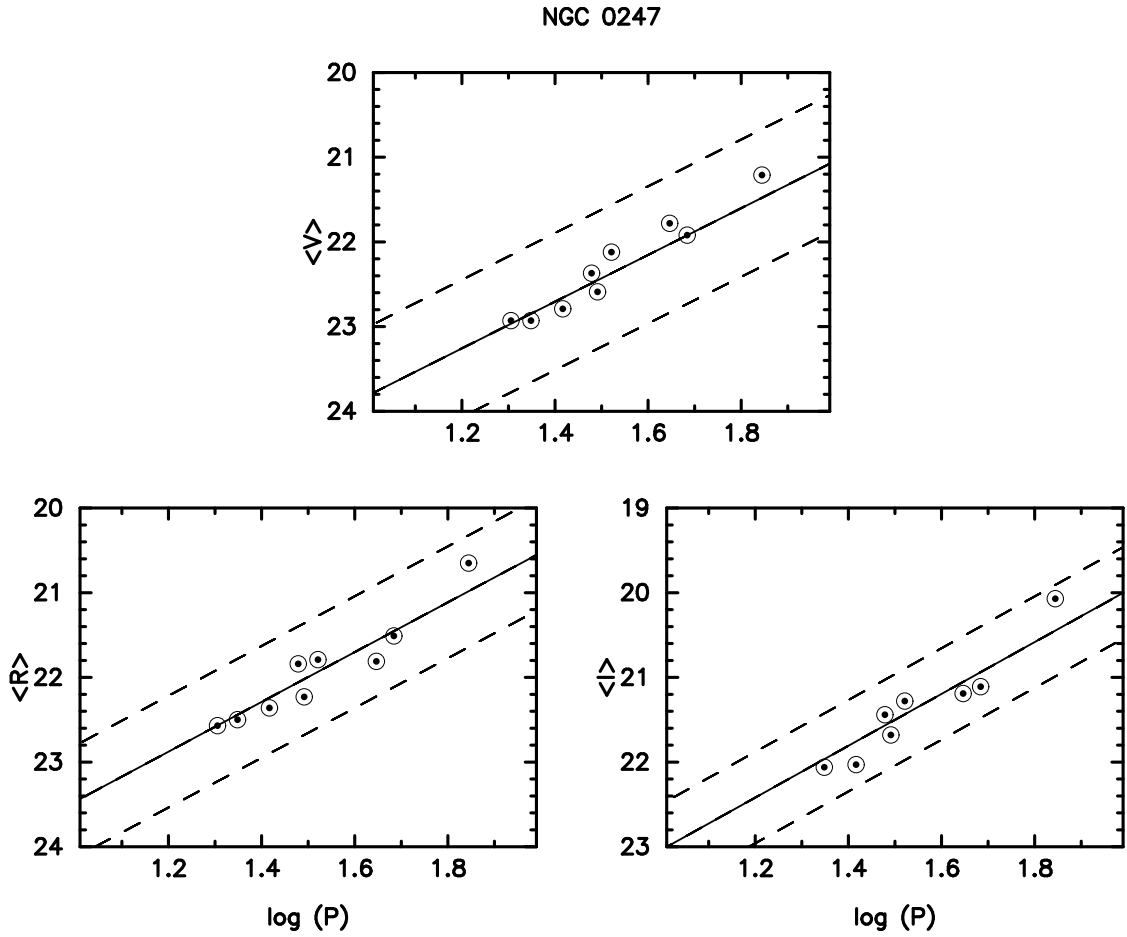}
\caption{Fits of the NGC~0247 Cepheid PL-relations in $V$, $R$, and
$I$ to the LMC PL-relation. Solid lines show the least-squares fit,
flanked by $\pm$2-sigma boundaries to the instability strip as derived from
LMC calibrators.}
\end{center}
\end{figure}

\subsection{PL Relations and Apparent Distance Moduli}

To determine apparent $VRI$ distance moduli, residuals about the PL
relations for NGC~0247 Cepheids were minimized relative to the mean
LMC PL relations given in \cite{mad91}, and updated to the VI
calibration of Udalski (2000).  For a given bandpass, the LMC~PL
relation was iteratively shifted relative to the NGC~0247 PL relation
until the $\chi^2$ of the fit was minimized.  The off-set determined
in this way is then the apparent distance modulus (for that bandpass)
of NGC~0247 with respect to the LMC.  The results of the PL-fits are
shown in Figure 3.  In the absence of other physical effects,
determination of the true distance modulus and reddening can obtained
by fitting the apparent moduli in different filters to an interstellar
extinction law (e.g., Cardelli et al. 1989)\footnote{Here we use A$_V$
= 3.2$\times$E(B-V) and A$_V$ = 2.45$\times$E(V-I). [GV08] choose to use
a slightly different reddening law, taken from Schlegel et al. (1998), giving
A$_V$ = 2.50$\times$E(V-I) which is only 2\% different from our
adopted value.} originally discussed in Freedman (1998).

In Figure 4, the apparent distance moduli at $VRI$ for the Cepheid
sample in NGC~0247 are plotted with respect to inverse wavelength.
The solid line gives a fit to a standard \cite{car89} Galactic
extinction law flanked by one-sigma error curves (dashed lines).  The
$VRI$ data are very well-fitted by an extinction curve (with a small
positive reddening equivalent to E(V-I) = 0.07~mag)\footnote{All
reddenings in this paper are given in terms of E(V-I). For those
wishing the E(B-V) equivalent the appropriate conversion factor is
E(B-V)/E(V-I) = 2.45/3.20 = 0.77} having an intercept corresponding to
a true distance modulus of $\mu_o$ = 27.81$\pm$0.10~mag
(3.65$\pm$0.16~Mpc). The solution using only $V$ and $I$ gives
essentially the same numbers ($\mu_o$ = 27.79$\pm$0.13~mag;
3.61$\pm$0.23~Mpc).

\begin{figure}
\begin{center}
\includegraphics[angle=-90, scale=0.60]{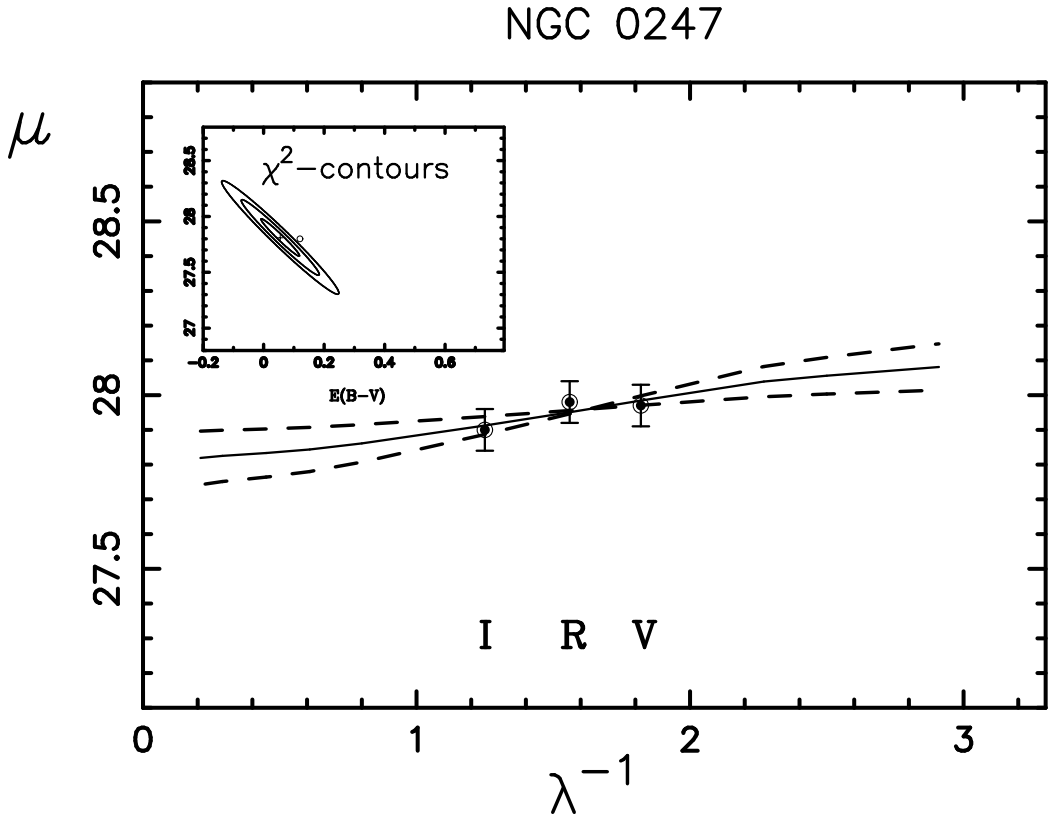}
\caption{Fit of the apparent distance moduli in $V$, $R$, and $I$ to a
Galactic extinction law (solid line). One-sigma errors on the fit are
shown with broken lines. Plotted contours are 2, 4 and 6 sigma.}
\end{center}
\end{figure}

\section{Comparision with Garcia-Varela et al. (2008)}

We have made a positional cross-correlation of our Cepheids with those
discovered by [GV08]. Six of our nine variables were recovered by the
Araucania Project, and the correct identification of these stars
across the two studies is reinforced by the independently derived
periods which agree to better than 10\% in most cases. We now combine
the two datasets, revise the periods when posible and update the $V$
and $I$ intensity-mean magnitudes. The results of that update are
given in Table 15. The combined lightcurves are shown in Figure 5.

\begin{figure}
\resizebox{0.31\hsize}{!}{\includegraphics[width=0.4\textwidth, angle=-90]{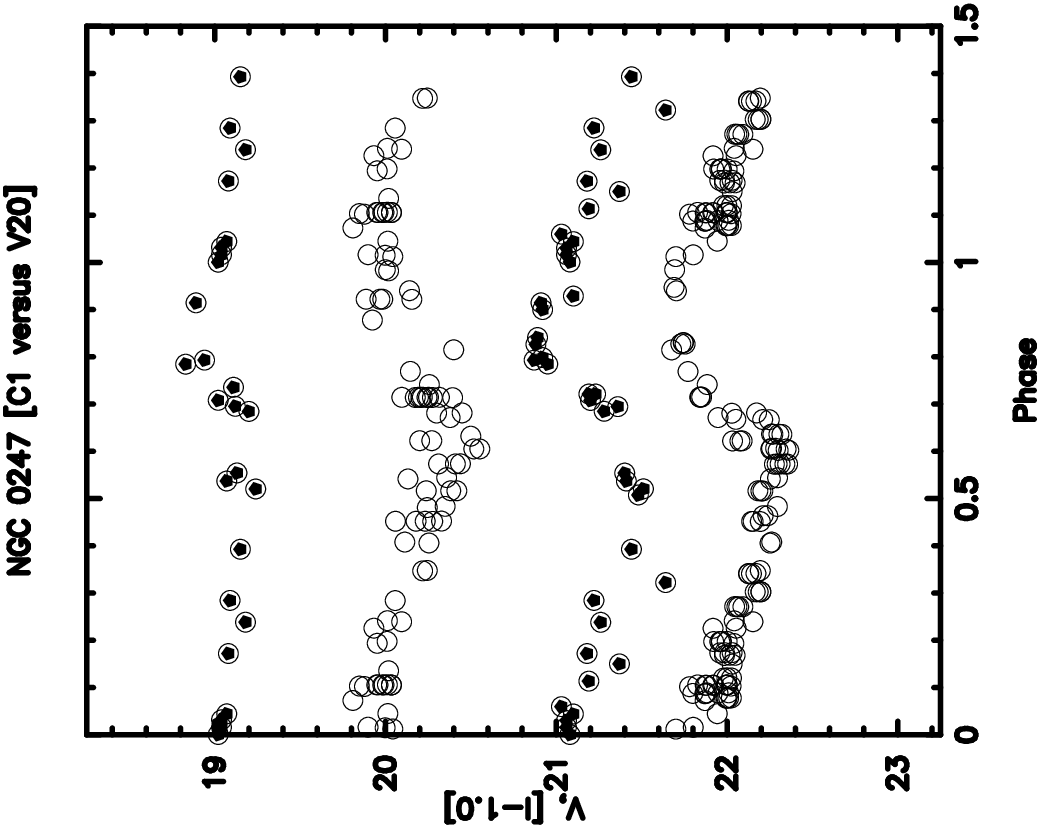}}\hspace{0.02cm}
\resizebox{0.31\hsize}{!}{\includegraphics[angle=-90]{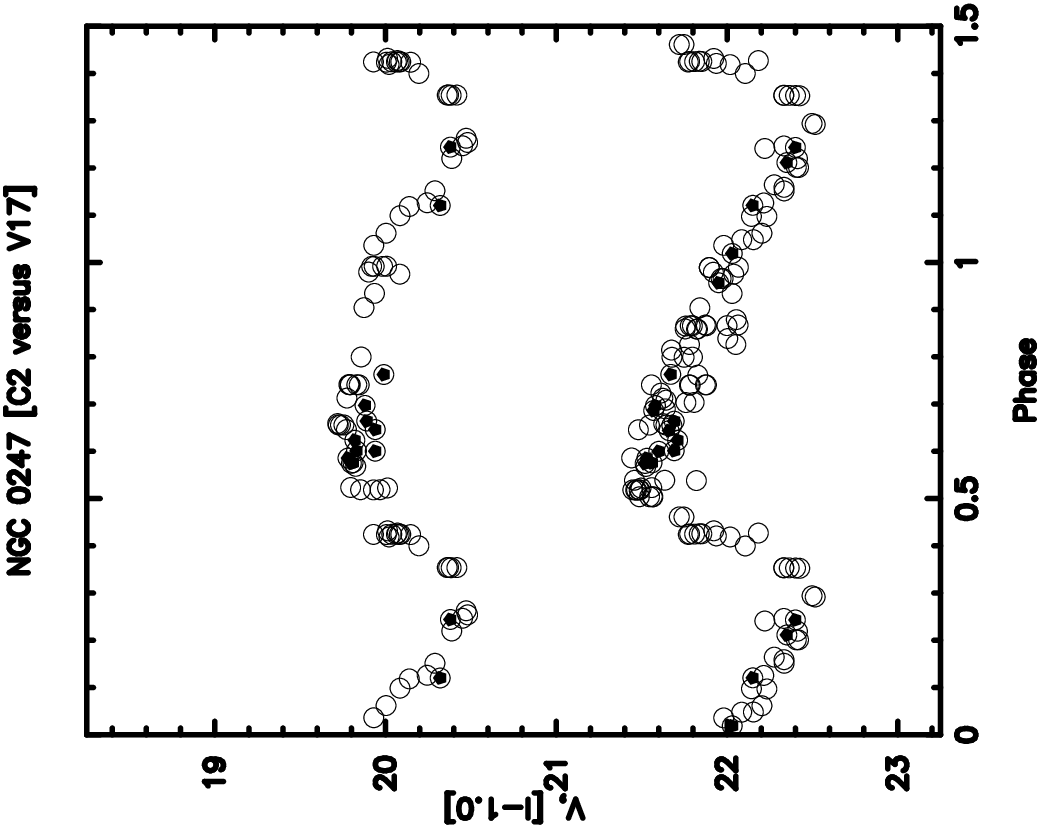}}
\resizebox{0.31\hsize}{!}{\includegraphics[angle=-90]{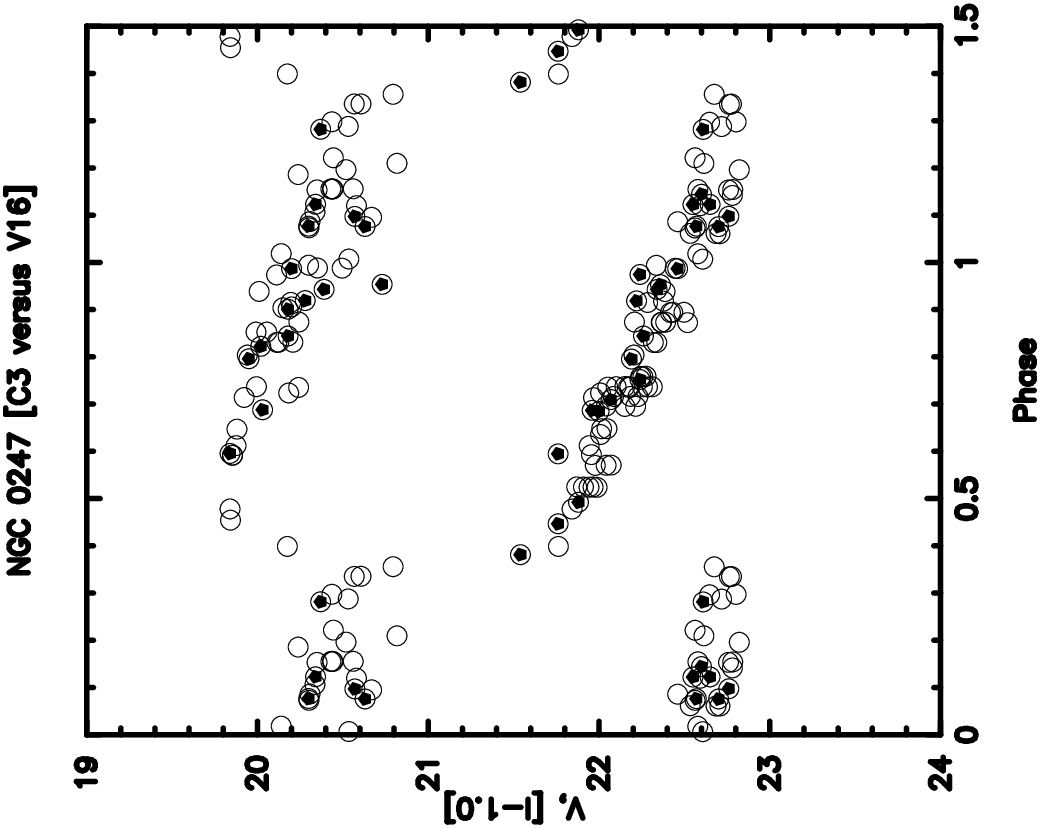}}\\

\resizebox{0.31\hsize}{!}{\includegraphics[width=0.4\textwidth, angle=-90]{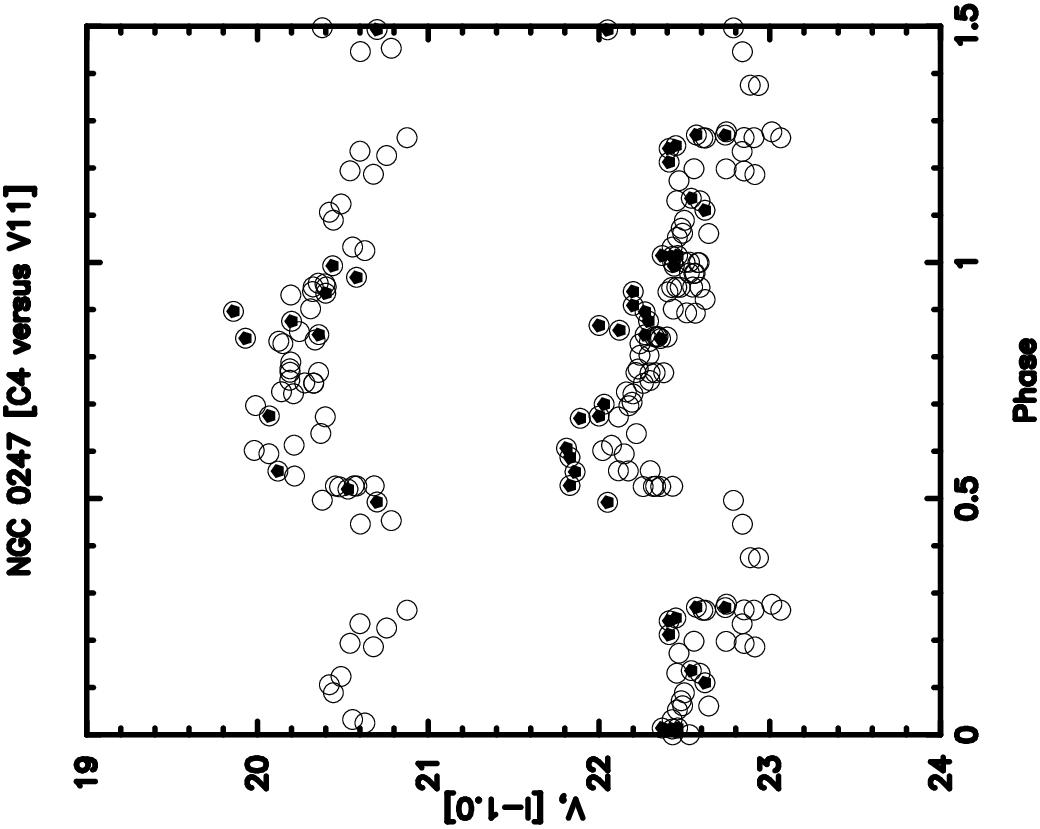}}\hspace{0.02cm}
\resizebox{0.31\hsize}{!}{\includegraphics[angle=-90]{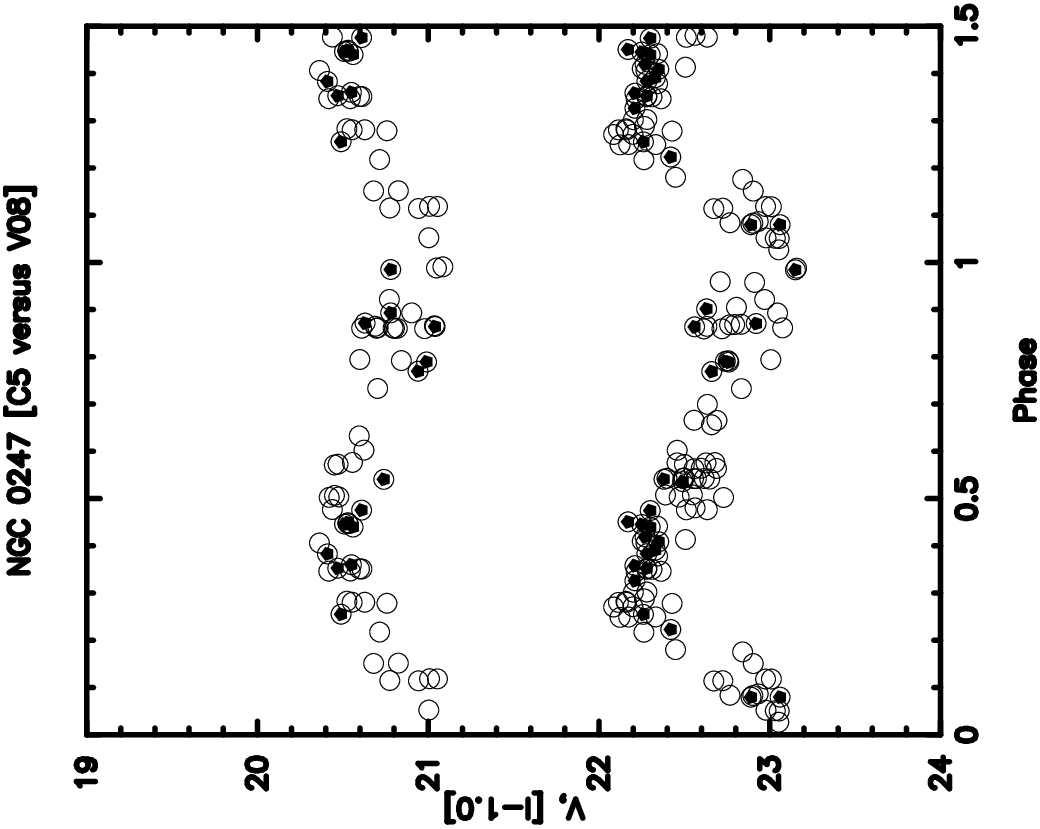}}
\resizebox{0.31\hsize}{!}{\includegraphics[angle=-90]{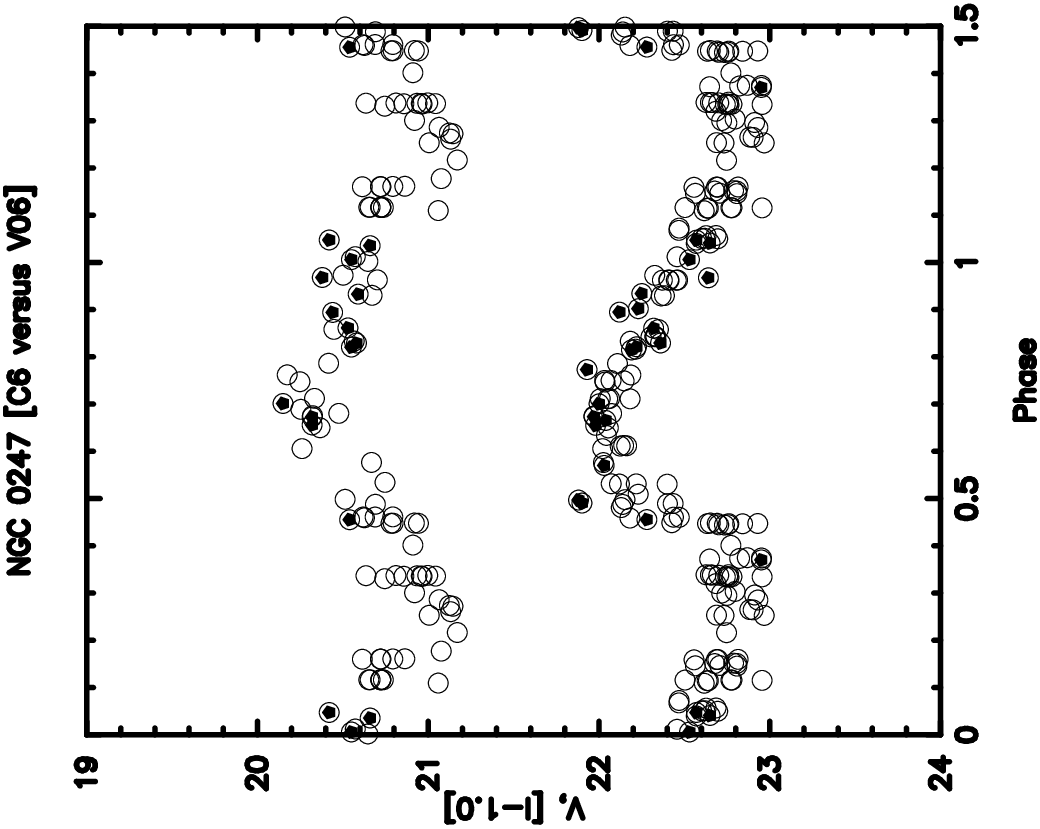}}\\

\caption{Combined $BVRI$ lightcurves for the individual Cepheids in
NGC~0247.  The plotted magnitude range is 5~mag in all
cases. Magnitude offsets, applied, to make the lightcurves
individually more visible, are given in the vertical-axis labels. In
order from top to bottom the lightcurves are $I$, $R$, $V$ and $B$. 
Solid points are from this paper; open circles are from [GV08].}
\end{figure}

The updated VRI [MF09] sample alone gives $\mu_V$ =
27.97$\pm$0.05~mag, $\mu_R$ = 27.98$\pm$0.06~mag, $\mu_I$ =
27.90$\pm$0.06~mag, E(V-I) = 0.11$\pm$0.03~mag resulting in $\mu_o$ =
27.81$\pm$0.05~mag or 3.65$\pm$0.08~Mpc for the 3-band fit, and
$\mu_o$ = 27.79$\pm$0.13~mag (3.61$\pm$0.23~Mpc) with E(V-I) =
0.07$\pm$0.04~mag for the VI fit alone.

We consider a progressive merger of the two datasets.  We first apply
our standard fitting techniques to the [GV08] preferred subset of 17
Cepheids, omitting as they did, the longest and shortest-period
Cepheids in their sample. We get $\mu_V$ = 28.21$\pm$0.05~mag,
$\mu_I$ = 28.05$\pm$0.06~mag, E(V-I) = 0.15$\pm$0.03~mag
resulting in $\mu_o$ = 27.82$\pm$0.08~mag or 3.66$\pm$0.14~Mpc.
This differs from the [GV08] solution by  +0.025~mag in the true
modulus.  

If we now update the [GV08] sample with the revised periods and
magnitudes for NGC~0247:[MF09]~C2 through C6 we get $\mu_V$ =
28.15$\pm$0.06~mag, $\mu_I$ = 28.03$\pm$0.06~mag, E(V-I) = 0.11$\pm$
0.03~mag resulting in $\mu_o$ = 27.87$\pm$0.09~mag or
3.75$\pm$0.15~Mpc.  Augmenting the [GV08] sample with
NGC~0247:[MF09]~C7 \& NGC~0247:[MF09]~C8, plus reintroducing
NGC~0247:[MF09]~C1 with its first-epoch period and magnitude, in
addition to its evolved values from [GV08] as described in Section 7
(below), we get $\mu_V$ = 28.13$\pm$0.05~mag, $\mu_I$ = 28.01$\pm$
0.06~mag, $E(V-I)$ = 0.11$\pm$0.03~mag resulting in $\mu_o$ =
27.85$\pm$ 0.09~mag or 3.72$\pm$0.15~Mpc.  The above results are
summarized in Table 16.

\section{The 70-Day Cepheid NGC~0247:[MF09]~C1}

In an attempt to update the period and combine the photometry for the
longest-period Cepheid in our sample, NGC~0247:[MF09]~C1, we quickly
found that the mean magnitudes and colors derived from our data did
not correspond to data published for it in the [GV08] study. Figure 6
shows the differences. In that plot our data are shown as circled
solid symbols, phased to our adopted period of 69.9 days.  Below those
light curves are the data from [GV08], shown as open cicles, phased to
their period of 65.862 days (with an arbitrarily added phase shift of
0.6 to align the lightcurves for ease of visual comparision). The
time-averaged V magnitudes differ by 0.8~mag, with the most recent
epoch being fainter; while the (V-I) colors differ by 0.23~mag, with
the most recent data being bluer. The sense of the change eliminates a
self-shrouding event as the possible cause. A remaining explanation
is that the structure of the star itself may have systematically
changed in the intervening quarter century: in the face of a rising
surface temperature (indicated by the decrease in the (V-I) color) and
the resulting increased surface brightness, the overall radius of this
star may have decreased significantly.  In the process the period
dropped by 6\%, from 70 to 66 days.

\begin{figure}
\begin{center}
\includegraphics[angle=-90, scale=0.85]{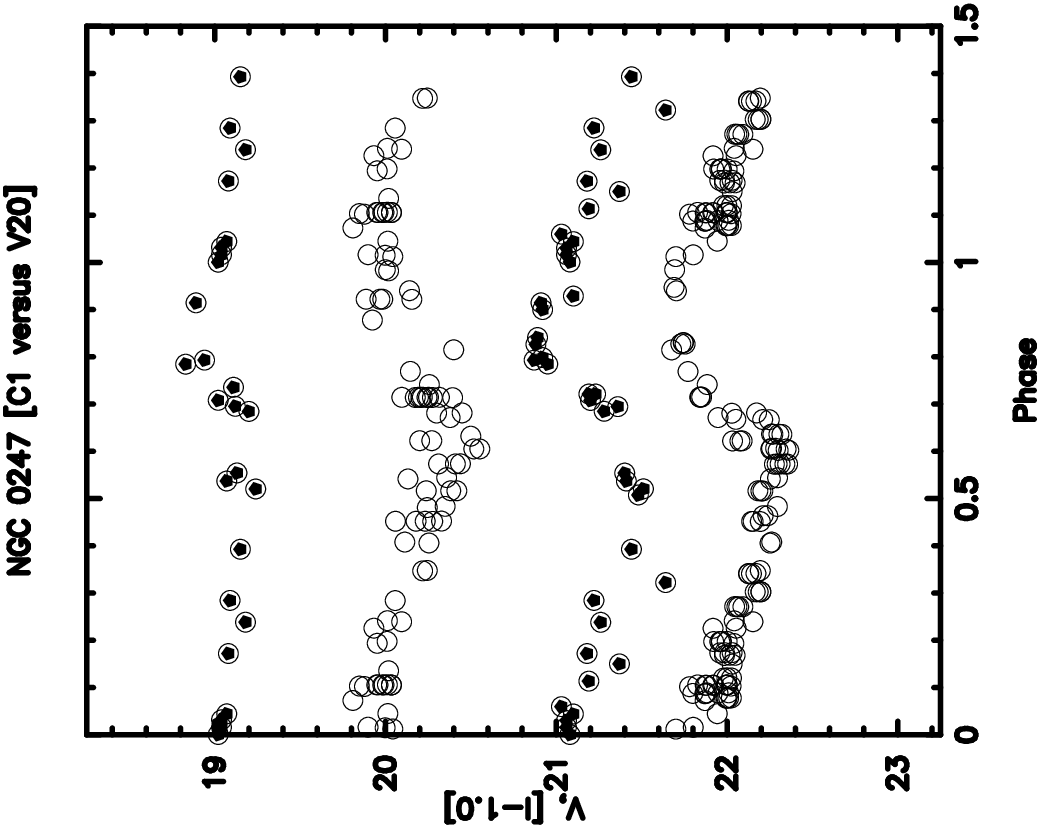}
\caption{The V and I lightcurves for the Cepheid C1 at the two epochs
surveyed by [MF09] (circled solid symbols) and [GV08] (open
circles). The I-band light curves are both displaced by one magnitude
upward in the figure for ease of viewing. In addition the earlier data
are displaced in phase by 0.6 cycles so as to align the two datasets
around maximum light.  The data are phased to a period of 69.9 days
for the [MF09] observations, and to a period of 65.862 days for the
[GV08] observations. The vertical displacement of the pairs of light
curves in V and in I is real, indicating that the star faded by nearly
0.8~mag between the times of the two studies.}
\end{center}
\end{figure}

A simple linear decrease of the period with time ($\Delta P/P$ =
0.0075 day/day) fails to phase the lightcurves over the total baseline
(and, in fact, destroys coherence within the individual observing
campaigns). Without undertaking more sophisticated modelling we
default to the next simplest conclusion that the period change was a
discontinuous event. Further monitoring of this star could be reveal
interesting aspects of the structure of Cepheids in general if this
behavior persists.

\section{Summary and Conclusions}

Nine Cepheids have been identified in the Galaxy NGC~0247. Six of
these variable stars have been independently found by [GV08].

The period and magnitude-updated VRI [MF09] sample alone gives
apparent moduli of $\mu_V$ = 27.97$\pm$0.05~mag, $\mu_R$ =
27.98$\pm$0.06~mag, $\mu_I$ = 27.90$\pm$0.06~mag, and $E(V-I)$ =
0.11$\pm$0.03~mag resulting in $\mu_o$ = 27.81$\pm$0.05~mag or
3.65$\pm$0.08~Mpc for the 3-band fit. These data yield a true distance
modulus of $\mu_{0} = 27.70\pm0.11$~mag corresponding to a metric
distance of 3.47$\pm$0.18~Mpc.

Combining our observations with newly published data from [GV08] in the V \& I
bands, and updating the periods accordingly, results in a reddening of
E(V-I) = 0.06$\pm$0.04~mag and a (preferred) true modulus of $\mu_{0}
= 27.81\pm0.05$~mag (3.65$\pm$0.08~Mpc). This is (fortuitously)
identical to the TRGB distance modulus recently published by
Karachentsev et al. (2006) further re-inforcing the consistency of
these two distance scales, which are based on largely independent
assumptions, and have very different systematics.

The 70-day Cepheid NGC~0247:[MF09]~C1 deserves follow-up observations
to see if the extraordinary changes in its magnitude, period and color
found between these epochs (first 1984-1992 and then 2002-2005) is an
on-going phenomenon.

\acknowledgements During the initial period in which these
observations were made WLF's research was supported in part by NSF
Grants AST 87-13889 and 9116496 on the extragalactic distance scale.
We thank Bob Williams who provided us with Director's Discretion Time
in 1988, and Irwin Horowitz who particiated in the early stages of
reducing the Las Campanas data.  We also thank Jose Garcia-Varela,
Grzegorz Pietrzynski and Wolfgang Gieren for providing their more
recently acquired Cepheid data in advance of publication. This
research has made use of the NASA/IPAC Extragalactic Database (NED)
which is operated by the Jet Propulsion Laboratory, Caltech, under
contract with the National Aeronautics and Space Administration.

\clearpage

\begin{deluxetable}{ccccclcccl}
\tablenum{1}
\tablewidth{5.5in}        
\tablecaption{Central Coordinates of Fields Observed in NGC~0247\label{tbl1}}
\tablehead{
\colhead{Field} & \colhead{} & \colhead{} & \colhead{} & \colhead{} & \colhead{RA(1950.0)} 
 & \colhead{}  & \colhead{} & \colhead{} & \colhead{Dec(1950.0)}  }
\startdata
NGC~0247:F1 &  &   &  &   & $00^{h}44^{m}35.4^{s}$ &  &  &  & $-20\arcdeg55\arcmin55\arcsec$ \\
NGC~0247:F2 &  &   &  &   & $00^{h}44^{m}40.9^{s}$ &  &  &  & $-21\arcdeg04\arcmin15\arcsec$ \\
NGC~0247:F3 &  &   &  &   & $00^{h}44^{m}33.4^{s}$ &  &  &  & $-20\arcdeg53\arcmin11\arcsec$ \\
\enddata
\end{deluxetable}

\begin{deluxetable}{llllll}
\tablenum{2}
\tablewidth{0pc}        
\tablecaption{Journal of Observations of NGC~0247 Cepheid Fields\label{tbl2}}
\tablehead{
\colhead{Date(UT)}      & \colhead{Telescope}   &
\colhead{Chip}          & \colhead{Scale}       &
\colhead{Fields}        & \colhead{Filters}}
\startdata
Nov. 24, 1984       &       CTIO    &       RCA     &      0.60~arcsec/pxl   &       1       &       $R$ \\
Sept. 19, 1985      &       CTIO    &       RCA     &      0.60~arcsec/pxl   &       1       &       $VRI$ \\
Sept. 20, 1985      &       CTIO    &       RCA     &      0.60~arcsec/pxl   &       1,2     &       $VR; VR$ \\
Sept. 21, 1985      &       CTIO    &       RCA     &      0.60~arcsec/pxl   &       2       &       $BVI$ \\
Sept. 22, 1985      &       CTIO    &       RCA     &      0.60~arcsec/pxl   &       1,2     &       $V; V$ \\
Sept. 23, 1985      &       CTIO    &       RCA     &      0.60~arcsec/pxl   &       1,2     &       $V; V$ \\
Dec. 06, 1985       &       CTIO    &       RCA     &      0.60~arcsec/pxl   &       1       &       $VR $ \\
Dec. 07, 1985       &       CTIO    &       RCA     &      0.60~arcsec/pxl   &       1,2     &       $BVI; BVI$ \\
Dec. 08, 1985       &       CTIO    &       RCA     &      0.60~arcsec/pxl   &       1,2     &       $VR; VR$ \\
Oct. 25, 1986       &       CTIO    &       RCA     &      0.60~arcsec/pxl   &       1,2     &       $BV; RI$ \\
Oct. 26, 1986       &       CTIO    &       RCA     &      0.60~arcsec/pxl   &       1,2,3   &       $VRI; BVRI; VRI$ \\
Nov. 08, 1986       &       CTIO    &       RCA     &      0.60~arcsec/pxl   &       1,2     &       $BV; BV$ \\
Nov. 09, 1986       &       CTIO    &       RCA     &      0.60~arcsec/pxl   &       1,2,3   &       $VRI; VRI; BVRI$ \\
Sept. 24, 1987      &       CTIO    &       RCA     &      0.60~arcsec/pxl   &       1,2,3   &       $BVRI; BV; BV$ \\
Oct. 13, 1987       &       CTIO    &       RCA     &      0.60~arcsec/pxl   &       1,2,3   &       $B; BVRI; BVRI$ \\
Oct. 22, 1987       &       CTIO    &       RCA     &      0.60~arcsec/pxl   &       1,2     &       $BVRI; BVI$ \\
Nov. 21, 1987       &       CTIO    &       RCA     &      0.60~arcsec/pxl   &       1,2     &       $BV; B, V$ \\
Nov. 25, 1987       &       CTIO    &       RCA     &      0.60~arcsec/pxl   &       1,2,3   &       $VRI; VRI; VRI$ \\
Sept. 10, 1988      &       CTIO    &       RCA     &      0.60~arcsec/pxl   &       1,2,3   &       $BVRI; BVRI; BV$ \\
Sept. 15, 1988      &       CTIO    &       RCA     &      0.60~arcsec/pxl   &       1,2,3   &       $BVI; BV; BV$ \\
Oct. 07, 1988       &       CTIO    &       RCA     &      0.60~arcsec/pxl   &       1,2,3   &       $BVRI; VRI; BVRI$ \\
Oct. 13, 1988       &       CTIO    &       RCA     &      0.60~arcsec/pxl   &       1,2,3   &       $VRI; BVRI; VRI$ \\
Nov. 07, 1988       &       CTIO    &       RCA     &      0.60~arcsec/pxl   &       1,2,3   &       $BV; BV; BV$ \\
Sept. 10, 1991      &       LCO     &       FORD1   &      0.16~arcsec/pxl   &       1,2,3   &       $R; R; VR$ \\
Sept. 11, 1991      &       LCO     &       FORD1   &      0.16~arcsec/pxl   &       1,2,3   &       $BVRI; BVRI; BVRI$ \\
Sept. 12, 1991      &       LCO     &       FORD1   &      0.16~arcsec/pxl   &       1,2,3   &       $BVRI; BVRI; BVRI$ \\
Sept. 13, 1991      &       LCO     &       FORD1   &      0.16~arcsec/pxl   &       1,2,3   &       $BVRI; BVRI; BVRI$ \\
Sept. 14, 1991      &       LCO     &       FORD1   &      0.16~arcsec/pxl   &       1,2,3   &       $BVRI; BVRI; BVRI$ \\
Oct. 02, 1992       &       LCO     &       TEK4    &      0.26~arcsec/pxl   &       1,2,3   &       $BVRI; BVRI; BVRI$ \\
Oct. 03, 1992       &       LCO     &       TEK4    &      0.26~arcsec/pxl   &       1,2,3   &       $BVRI; BVRI; VRI$ \\
Nov. 20, 1992       &       LCO     &       TEK4    &      0.26~arcsec/pxl   &       1,2     &       $BVRI; RI$ \\
Nov. 21, 1992       &       LCO     &       TEK4    &      0.26~arcsec/pxl   &       2,3     &       $BVR; BVRI$ \\
Dec. 20, 1992       &       LCO     &       TEK3    &      0.23~arcsec/pxl   &       3       &       $BVRI$ \\
Dec. 21, 1992       &       LCO     &       TEK3    &      0.23~arcsec/pxl   &       2       &       $BVRI$ \\
Dec. 22, 1992       &       LCO     &       TEK3    &      0.23~arcsec/pxl   &       1,3     &       $VRI; BVRI$ \\
\enddata
\end{deluxetable}

\begin{deluxetable}{llcc}
\tablenum{3}
\tablewidth{0pc}        
\tablecaption{Positions for [MF09] Cepheids \label{tbl3}}
\tablehead{
\colhead{Name} & \colhead{P(days)}  & \colhead{RA~(2000)}   & \colhead{DEC~(2000)}     }
\startdata
NGC~0247:[MF09]~C1     & 65.86  & 00:47:10.6  & -20:40:11     \\
NGC~0247:[MF09]~C2     & 48.53  & 00:47:03.8  & -20:41:04     \\
NGC~0247:[MF09]~C3     & 48.38  & 00:47:10.5  & -20:47:21     \\
NGC~0247:[MF09]~C4     & 33.23  & 00:47:10.1  & -20:48:45     \\
NGC~0247:[MF09]~C5     & 30.931 & 00:47:03.5  & -20:47:59     \\
NGC~0247:[MF09]~C6     & 27.785 & 00:47:07.6  & -20:37:51     \\
NGC~0247:[MF09]~C7     & 26.2   & 00:47:01.1  & -20:39:12     \\
NGC~0247:[MF09]~C8     & 22.3   & 00:47:01.1  & -20:37:58     \\
NGC~0247:[MF09]~C9     & 20.2   & 00:47:01.4  & -20:39:55     \\
\enddata
\end{deluxetable}

\begin{deluxetable}{cclclcl}
\tablecolumns{7}
\tablenum{4}
\tablewidth{0pt}
\tablecaption{Observations of C1 (P = 69.9 days) \label{tbl4}}
\tablehead{
\colhead{Filter} & \colhead{}  & \colhead{Mag.} & \colhead{} & \colhead{Error}
& \colhead{} &  \colhead{Julian Day}\\
}
\startdata
$B$ &  &  &  &  &  &   \\
  \  & &    22.23 & &    0.03 & &    2447441.6878\nl
  \  & &    21.82 & &    0.02 & &    2446407.6021\\
   \  & &   22.30 & &    0.04 & &    2447120.5347\\
    \  & &   22.45 & &   0.04 & &    2447472.5924\\
    \  & &   22.23 & &    0.04 & &    2447090.5611\\
    \  & &   22.66 & &    0.05 & &    2446728.6868\\
    \  & &   22.71 & &    0.10 & &  2447414.7347\\
    \  & &   22.57 & &    0.09 & &    2447419.6694\\
    \  & &   22.71 & &    0.08 & &    2447062.6007\\
    \  & &   22.24 & &    0.06 & &    2446743.6097\\
    \  & &   22.03 & &    0.04 & &    2448510.6653\\
    \  & &   21.98 & &    0.02 & &    2448511.7306\\
    \  & &   22.01 & &    0.03 & &    2448512.7177\\
    \  & &   22.04 & &    0.03 & &    2448513.7368\\
    \  & &   22.46 & &    0.09 & &    2448946.6891\\
    \  & &   22.39 & &    0.09 & &    2448898.8180\\
$V$ &  &  &  &  &  &   \\
 \  & &   21.20 & &    0.02 & &    2447441.6847\\
 \  & &   20.91 & &    0.02 & &    2446407.6208\\
 \  & &   21.19 & &    0.03 & &    2447120.5403\\
\  & &   21.37 & &    0.03 & &    2447472.5979\\
\  & &   21.28 & &    0.03 & &    2447090.5708\\
\  & &   21.48 & &    0.04 & &    2446728.6819\\
\  & &   21.64 & &    0.05 & &    2447414.7306\\
\  & &   21.44 & &    0.04 & &    2447419.6653\\
\  & &   21.22 & &    0.03 & &    2447062.5917\\
\  & &   21.23 & &    0.04 & &    2446743.6229\\
  \  & &   21.08 & &    0.03 & &    2448510.6576\\
  \  & &   21.06 & &    0.02 & &    2448511.7323\\
  \  & &   21.06 & &    0.02 & &    2448512.7198\\
  \  & &   21.10 & &    0.02 & &    2448513.6840\\
 \  & &   21.26 & &    0.02 & &    2448946.6490\\
 \  & &   21.40 & &    0.03 & &    2448898.8294\\
  \  & &   21.41 & &    0.03 & &    2448897.6504\\
  \  & &   21.03 & &    0.06 & &    2448514.7785\\
  \  & &   20.92 & &    0.02 & &    2446406.6125\\
  \  & &   20.10 & &    0.02 & &    2446408.6236\\
   \  & &   21.18 & &   0.03 & &    2447124.6278\\
   \  & &   21.19 & &    0.04 & &    2446743.6250\\
   \  & &   21.51 & &    0.03 & &    2446729.5938\\
   \  & &   20.95 & &    0.03 & &    2446328.6632\\
   \  & &   20.92 & &    0.02 & &    2446329.6243\\
   \  & &   20.88 & &    0.03 & &    2446331.5681\\
   \  & &   20.89 & &    0.04 & &    2446332.5826\\
   \  & &   20.87 & &    0.02 & &    2447447.6328\\
   \  & &   21.36 & &    0.05 & &    2448978.5743\\
$R$ &  &  &  &  &  &   \\
   \  & &   20.60 & &    0.01 & &    2447441.6910\\
   \  & &   20.72 & &    0.03 & &    2447090.5514\\
   \  & &   20.75 & &    0.02 & &    2447414.7417\\
   \  & &   20.58 & &    0.03 & &    2447062.5854\\
   \  & &   20.57 & &    0.03 & &    2448510.6493\\
  \  & &   20.60 & &    0.01 & &    2448511.7340\\
\  & &   20.55 & &    0.02 & &    2448512.7198\\
  \  & &   20.55 & &    0.02 & &    2448513.6924\\
  \  & &   20.79 & &    0.01 & &    2448946.6623\\
  \  & &   20.79 & &    0.02 & &    2448897.6596\\
  \  & &   20.81 & &    0.02 & &    2448898.8385\\
  \  & &   20.57 & &    0.02 & &    2448509.8417\\
  \  & &   20.75 & &    0.02 & &    2446029.5344\\
  \  & &   20.40 & &    0.02 & &    2446406.6069\\
  \  & &   20.44 & &    0.01 & &    2446408.6146\\
  \  & &   20.65 & &    0.02 & &    2447124.6236\\
  \  & &   20.66 & &    0.02 & &    2446744.6194\\
  \  & &   20.91 & &    0.02 & &    2446729.5833\\
  \  & &   20.42 & &    0.02 & &    2446328.6771\\
  \  & &   20.40 & &    0.02 & &    2446329.6389\\
  \  & &   20.45 & &    0.01 & &    2447447.6472\\
  \  & &   20.73 & &    0.04 & &    2448978.6041\\
$I$ &  &  &  &  &  &   \\
  \  & &   20.02 & &    0.02 & &    2447441.6965\\
  \  & &   20.20 & &    0.04 & &    2447090.5556\\
  \  & &   20.09 & &    0.08 & &    2447062.5819\\
  \  & &   20.02 & &    0.03 & &    2448510.6410\\
  \  & &   20.04 & &    0.02 & &    2448511.7337\\
  \  & &   20.04 & &    0.03 & &    2448512.8333\\
  \  & &   20.07 & &    0.03 & &    2448513.7007\\
 \  & &   20.18 & &    0.02 & &    2448946.6748\\
   \  & &   20.07 & &    0.02 & &    2448897.6686\\
   \  & &   20.13 & &    0.03 & &    2448898.8482\\
   \  & &    19.89 & &   0.03 & &    2446407.6104\\
  \  & &   20.08 & &    0.05 & &    2447124.6375\\
  \  & &   20.11 & &    0.04 & &    2446744.6153\\
  \  & &   20.24 & &    0.04 & &    2446729.5875\\
  \  & &   20.15 & &    0.03 & &    2447419.6590\\
   \  & &    19.83 & &   0.05 & &    2446328.6528\\
   \  & &    19.94 & &   0.02 & &    2447447.6528\\
  \  & &   20.12 & &    0.04 & &    2448978.6166\\
\enddata
\end{deluxetable}

\begin{deluxetable}{cclclcl}
\tablecolumns{7}
\tablenum{5}
\tablewidth{0pt}
\tablecaption{Observations of C2 (P = 48.3 days) \label{tbl5}}
\tablehead{
\colhead{Filter} & \colhead{}  & \colhead{Mag.} & \colhead{} & \colhead{Error}
& \colhead{} &  \colhead{Julian Day}\\
}
\startdata
$B$ &  &  &  &  &  &   \\
  \  & &   22.19 & &    0.03 & &    2447441.6878\\
  \  & &   23.00 & &    0.09 & &    2447120.5347\\
  \  & &   23.37 & &    0.11 & &    2447472.5924\\
  \  & &   23.34 & &    0.12 & &    2447419.6694\\
  \  & &   22.42 & &    0.07 & &    2447062.6007\\
  \  & &   22.49 & &    0.07 & &    2448510.6653\\
  \  & &   22.35 & &    0.04 & &    2448511.7306\\
  \  & &   22.47 & &    0.04 & &    2448512.7177\\
  \  & &   22.38 & &    0.04 & &    2448513.7368\\
  \  & &   22.20 & &    0.07 & &    2448946.6891\\
  \  & &   22.30 & &    0.09 & &    2448897.6395\\
  \  & &   22.26 & &    0.09 & &    2448898.8180\\
$V$ &  &  &  &  &  &    \\
  \  & &   21.52 & &    0.02 & &    2447441.6847\\
  \  & &   21.95 & &    0.05 & &    2447120.5403\\
  \  & &   22.35 & &    0.09 & &    2447472.5979\\
  \  & &   22.03 & &    0.09 & &    2447414.7306\\
  \  & &   22.15 & &    0.04 & &    2447419.6653\\
  \  & &   21.67 & &    0.06 & &    2447062.5917\\
  \  & &   21.69 & &    0.06 & &    2448510.6576\\
  \  & &   21.71 & &    0.04 & &    2448511.7323\\
  \  & &   21.66 & &    0.04 & &    2448512.7198\\
  \  & &   21.69 & &    0.05 & &    2448513.6840\\
  \  & &   21.53 & &    0.04 & &    2448946.6490\\
  \  & &   21.60 & &    0.03 & &    2448898.8294\\
  \  & &   21.56 & &    0.04 & &    2448897.6504\\
  \  & &   21.57 & &    0.10 & &    2448514.7785\\
  \  & &   21.58 & &    0.03 & &    2447447.6328\\
  \  & &   22.40 & &    0.12 & &    2448978.5743\\
$R$ &  &  &  &  &  &    \\
  \  & &   21.17 & &    0.02 & &    2447441.6910\\
  \  & &   21.14 & &    0.04 & &    2447062.5854\\
  \  & &   21.21 & &    0.04 & &    2448510.6493\\
  \  & &   21.29 & &    0.05 & &    2448511.7340\\
  \  & &   21.28 & &    0.03 & &    2448512.7198\\
  \  & &   21.32 & &    0.03 & &    2448513.6924\\
  \  & &   21.15 & &    0.02 & &    2448946.6623\\
  \  & &   21.24 & &    0.03 & &    2448897.6596\\
 \  & &   21.21 & &    0.04 & &    2448898.8385\\
  \  & &   21.25 & &    0.03 & &    2448509.8417\\
  \  & &   21.20 & &    0.03 & &    2447447.6472\\
  \  & &   21.74 & &    0.10 & &  2448978.6041\\
$I$ &  &  &  &  &  &    \\
  \  & &   20.80 & &    0.04 & &    2447441.6965\\
  \  & &   20.99 & &    0.14 & &    2447062.5819\\
  \  & &   20.83 & &    0.08 & &    2448510.6410\\
  \  & &   20.82 & &    0.05 & &    2448511.7337\\
  \  & &   20.94 & &    0.06 & &    2448512.8333\\
  \  & &   20.89 & &    0.06 & &    2448513.7007\\
  \  & &   20.78 & &    0.04 & &    2448946.6748\\
  \  & &   20.81 & &    0.04 & &    2448897.6686\\
  \  & &   20.94 & &    0.07 & &    2448898.8482\\
  \  & &   21.32 & &    0.07 & &    2447419.6590\\
  \  & &   20.88 & &    0.04 & &    2447447.6528\\
  \  & &   21.38 & &    0.13 & &    2448978.6166\\
\enddata
\end{deluxetable}

\begin{deluxetable}{cclclcl}
\tablecolumns{7}
\tablenum{6}
\tablewidth{0pt}
\tablecaption{Observations of C3 (P = 44.3 days) \label{tbl6}}
\tablehead{
\colhead{Filter} & \colhead{}  & \colhead{Mag.} & \colhead{} & \colhead{Error}
& \colhead{} &  \colhead{Julian Day}\\
}
\startdata
$B$ &  &  &  &  &  &    \\
  \  & &   23.72 & &   0.17 & &    2447447.6618\\
  \  & &   23.33 & &   0.20 & &    2448510.6931\\
  \  & &   23.64 & &   0.17 & &    2448511.6833\\
  \  & &   23.47 & &   0.19 & &    2448513.6597\\
  \  & &   22.58 & &   0.14 & &    2448977.5926\\
  \  & &   23.08 & &   0.14 & &    2448947.5667\\
  \  & &   23.41 & &   0.18 & &    2448897.6799\\
  \  & &   23.02 & &   0.06 & &    2446407.5903\\
  \  & &   23.23 & &   0.12 & &    2447120.5535\\
  \  & &   23.09 & &   0.15 & &    2446729.6028\\
  \  & &   22.17 & &   0.16 & &    2447062.6222\\
$V$ &  &  &  &  &  &   \\
  \  & &   22.46 & &   0.07 & &    2447441.6278\\
 \  & &   22.65 & &    0.13 & &    2447447.6688\\
 \  & &   22.70 & &    0.14 & &    2448510.7049\\
  \  & &   22.76 & &    0.12 & &    2448511.6750\\
 \  & &   22.55 & &    0.10 & &  2448512.7733\\
   \  & &   22.60 & &    0.10 & &  2448513.7504\\
  \  & &   21.88 & &    0.06 & &    2447419.6910\\
  \  & &   21.96 & &    0.04 & &    2446407.5799\\
  \  & &   22.07 & &    0.04 & &    2446408.5924\\
  \  & &   22.24 & &    0.06 & &    2447120.5597\\
  \  & &   22.26 & &    0.08 & &    2447124.6528\\
  \  & &   22.00 & &    0.12 & &    2447472.6056\\
  \  & &   22.61 & &    0.15 & &    2446744.6312\\
  \  & &   22.57 & &    0.12 & &    2447090.5771\\
  \  & &   22.34 & &    0.09 & &    2446729.5993\\
  \  & &   21.54 & &    0.10 & &  2447414.7715\\
  \  & &   22.36 & &    0.11 & &    2446330.6319\\
 \  & &   22.24 & &    0.16 & &    2446331.5819\\
   \  & &   21.76 & &    0.05 & &    2447062.6278\\
  \  & &   22.22 & &    0.05 & &    2448947.5213\\
  \  & &   21.76 & &    0.07 & &    2448977.5428\\
  \  & &   22.19 & &    0.08 & &    2448897.6872\\
$R$ &  &  &  &  &  &    \\
  \  & &   21.87 & &    0.05 & &    2447441.6340\\
  \  & &   22.02 & &    0.08 & &    2447447.6764\\
  \  & &   22.20 & &    0.08 & &    2448510.7125\\
  \  & &   22.20 & &    0.08 & &    2448511.6674\\
  \  & &   22.04 & &    0.06 & &    2448513.7535\\
  \  & &   21.65 & &    0.14 & &    2448977.5622\\
  \  & &   21.84 & &    0.03 & &    2448947.5392\\
  \  & &   21.68 & &    0.04 & &    2448897.6998\\
  \  & &   21.91 & &    0.05 & &    2446729.6097\\
  \  & &   22.00 & &    0.08 & &    2448512.6653\\
  \  & &   22.00 & &    0.08 & &    2448509.8576\\
  \  & &   21.55 & &    0.05 & &    2446406.5993\\
  \  & &   21.51 & &    0.04 & &    2446408.5833\\
  \  & &   21.75 & &    0.06 & &    2447124.6479\\
  \  & &   22.11 & &    0.10 & &    2446744.6368\\
  \  & &   21.72 & &    0.09 & &    2446329.6562\\
  \  & &   21.62 & &    0.09 & &    2447081.5653\\
  \  & &   21.68 & &    0.04 & &    2448946.7196\\
$I$ &  &  &  &  &  &    \\
  \  & &   21.20 & &    0.07 & &    2447441.6396\\
  \  & &   21.34 & &    0.09 & &    2447447.6819\\
  \  & &   21.63 & &    0.12 & &    2448510.7201\\
  \  & &   21.57 & &    0.09 & &    2448511.6583\\
  \  & &   20.84 & &    0.09 & &    2448977.5759\\
  \  & &   21.28 & &    0.05 & &    2448947.5525\\
  \  & &   20.95 & &    0.05 & &    2448897.7084\\
  \  & &   21.39 & &    0.12 & &    2446729.6139\\
  \  & &   21.03 & &    0.07 & &    2446407.6347\\
  \  & &   21.18 & &    0.10 & &  2447124.6438\\
  \  & &   21.37 & &    0.12 & &    2446744.6417\\
  \  & &   21.30 & &    0.11 & &    2447090.5882\\
  \  & &   21.73 & &    0.16 & &    2446330.6611\\
  \  & &   21.02 & &    0.11 & &    2448898.8788\\
  \  & &   21.18 & &    0.07 & &    2448946.7350\\
\enddata
\end{deluxetable}

\begin{deluxetable}{cclclcl}
\tablecolumns{7}
\tablenum{7}
\tablewidth{0pt}
\tablecaption{Observations of C4 (P = 33.2 days) \label{tbl7}}
\tablehead{
\colhead{Filter} & \colhead{}  & \colhead{Mag.} & \colhead{} & \colhead{Error}
& \colhead{} &  \colhead{Julian Day}\\
}
\startdata
$B$  &  &  &  &  &  &    \\
  \  & &   23.10 & &    0.09 & &    2447447.6618\\
  \  & &   23.29 & &   0.12 & &    2448510.6931\\
  \  & &   23.33 & &    0.10 & &  2448511.6833\\
  \  & &   23.62 & &   0.17 & &    2448513.6597\\
  \  & &   22.82 & &   0.15 & &    2448977.5926\\
  \  & &   23.34 & &   0.18 & &    2448947.5667\\
  \  & &   22.88 & &    0.10 & &  2448897.6799\\
  \  & &   22.04 & &   0.12 & &    2448898.8590\\
  \  & &   22.40 & &    0.05 & &    2446407.5903\\
  \  & &   23.19 & &   0.13 & &    2447120.5535\\
  \  & &   22.58 & &   0.11 & &    2446743.6340\\
  \  & &   22.60 & &    0.10 & &    2448898.6477\\
$V$ &  &  &  &  &  &    \\
  \  & &   22.00 & &    0.05 & &    2447441.6278\\
 \  & &   22.12 & &    0.06 & &    2447447.6688\\
 \  & &   22.27 & &    0.08 & &    2448510.7049\\
  \  & &   22.29 & &    0.09 & &    2448511.6750\\
  \  & &   22.20 & &    0.06 & &    2448512.7733\\
  \  & &   22.20 & &    0.08 & &    2448513.7504\\
 \  & &   22.37 & &    0.08 & &    2447419.6868\\
\  & &   22.36 & &   0.12 & &    2447081.5562\\
  \  & &   22.46 & &    0.08 & &    2447419.6910\\
  \  & &   21.86 & &    0.05 & &    2446407.5799\\
  \  & &   21.83 & &    0.06 & &    2446408.5924\\
  \  & &   22.43 & &    0.06 & &    2447120.5597\\
  \  & &   22.54 & &    0.08 & &    2447124.6528\\
  \  & &   21.81 & &    0.07 & &    2447472.6056\\
  \  & &   21.89 & &    0.07 & &    2446743.6403\\
  \  & &   22.03 & &    0.09 & &    2446744.6312\\
  \  & &   22.62 & &    0.10 & &  2447090.5771\\
  \  & &   22.45 & &    0.08 & &    2446729.5993\\
  \  & &   22.00 & &    0.09 & &    2447414.7715\\
  \  & &   22.41 & &    0.10 & &  2446329.6688\\
  \  & &   22.41 & &    0.08 & &    2446330.6319\\
  \  & &   22.57 & &   0.18 & &    2446331.5819\\
  \  & &   22.74 & &   0.12 & &    2447062.6278\\
  \  & &   22.44 & &    0.05 & &    2448947.5213\\
  \  & &   21.83 & &    0.07 & &    2448898.8663\\
  \  & &   22.27 & &   0.12 & &    2448977.5428\\
  \  & &   22.05 & &    0.06 & &    2448897.6872\\
$R$ &  &  &  &  &  &    \\
  \  & &   21.70 & &    0.09 & &    2447441.6340\\
  \  & &   21.94 & &    0.06 & &    2448510.7125\\
  \  & &   21.94 & &    0.08 & &    2448511.6674\\
  \  & &   21.83 & &    0.05 & &    2448513.7535\\
  \  & &   21.49 & &   0.12 & &    2448977.5622\\
  \  & &   22.12 & &    0.04 & &    2448947.5392\\
  \  & &   21.86 & &    0.04 & &    2448897.6998\\
  \  & &   21.74 & &    0.04 & &    2448898.6209\\
  \  & &   22.18 & &    0.09 & &    2446729.6097\\
  \  & &   21.78 & &    0.07 & &    2448512.6653\\
  \  & &   21.67 & &    0.05 & &    2448509.8576\\
  \  & &   21.69 & &    0.05 & &    2446406.5993\\
  \  & &   21.63 & &    0.07 & &    2446408.5833\\
  \  & &   21.90 & &    0.06 & &    2447124.6479\\
  \  & &   21.62 & &    0.06 & &    2446744.6368\\
  \  & &   21.70 & &    0.08 & &    2447081.5653\\
  \  & &   22.13 & &    0.06 & &    2448946.7196\\
$I$ &  &  &  &  &  &    \\
  \  & &   21.07 & &    0.07 & &    2447441.6396\\
  \  & &   21.36 & &    0.08 & &    2448510.7201\\
  \  & &   21.20 & &    0.08 & &    2448511.6583\\
  \  & &   21.40 & &    0.08 & &    2448513.6340\\
  \  & &   20.86 & &   0.11 & &    2448977.5759\\
  \  & &   21.44 & &    0.06 & &    2448947.5525\\
  \  & &   21.70 & &    0.10 & &  2448897.7084\\
  \  & &   21.53 & &    0.10 & &  2448898.6084\\
  \  & &   21.12 & &    0.08 & &    2446407.6347\\
  \  & &   20.93 & &   0.16 & &    2447081.5694\\
  \  & &   21.58 & &    0.09 & &    2448946.7350\\
\enddata
\end{deluxetable}

\begin{deluxetable}{cclclcl}
\tablecolumns{7}
\tablenum{8}
\tablewidth{0pt}
\tablecaption{Observations of C5 (P = 31.0 days) \label{tbl8}}
\tablehead{
\colhead{Filter} & \colhead{}  & \colhead{Mag.} & \colhead{} & \colhead{Error}
& \colhead{} &  \colhead{Julian Day}\\
}
\startdata
$B$  &  &  &  &  &  &    \\
  \  & &   23.45 & &   0.13 & &    2447447.6618\\
  \  & &   22.72 & &    0.08 & &    2448510.6931\\
  \  & &   22.56 & &    0.07 & &    2448511.6833\\
  \  & &   22.62 & &    0.06 & &    2448512.7618\\
  \  & &   22.95 & &   0.12 & &    2448513.6597\\
  \  & &   22.85 & &   0.20 & &    2448977.5926\\
  \  & &   23.08 & &   0.04 & &    2448947.5667\\
  \  & &   22.78 & &    0.06 & &    2446407.5903\\
  \  & &   23.10 & &    0.10 & &  2447120.5535\\
  \  & &   22.97 & &    0.10 & &  2447090.5819\\
  \  & &   23.53 & &   0.16 & &    2446729.6028\\
  \  & &   23.64 & &   0.12 & &    2447419.6812\\
  \  & &   22.81 & &   0.20 & &    2448898.6477\\
$V$ &  &  &  &  &  &    \\
  \  & &   22.76 & &    0.08 & &    2447441.6278\\
  \  & &   23.15 & &   0.12 & &    2447447.6688\\
  \  & &   22.28 & &    0.08 & &    2448510.7049\\
  \  & &   22.28 & &    0.08 & &    2448511.6750\\
  \  & &   22.27 & &    0.07 & &    2448512.7733\\
  \  & &   22.17 & &    0.06 & &    2448513.7504\\
  \  & &   22.89 & &    0.10 & &  2447419.6868\\
  \  & &   23.06 & &   0.16 & &    2447419.6910\\
  \  & &   22.21 & &    0.06 & &    2446406.5903\\
  \  & &   22.21 & &    0.05 & &    2446407.5799\\
  \  & &   22.33 & &    0.06 & &    2446408.5924\\
  \  & &   22.35 & &    0.06 & &    2447120.5597\\
  \  & &   22.38 & &    0.08 & &    2447124.6528\\
  \  & &   22.74 & &   0.18 & &    2447472.6056\\
  \  & &   22.42 & &   0.11 & &    2446743.6403\\
  \  & &   22.26 & &   0.12 & &    2446744.6312\\
  \  & &   22.30 & &    0.07 & &    2447090.5771\\
  \  & &   22.66 & &    0.09 & &    2446729.5993\\
  \  & &   22.92 & &   0.15 & &    2446330.6319\\
  \  & &   22.49 & &    0.09 & &    2447062.6278\\
  \  & &   22.30 & &    0.04 & &    2448947.5213\\
  \  & &   22.63 & &   0.13 & &    2448898.8663\\
  \  & &   22.25 & &   0.11 & &    2448977.5428\\
  \  & &   22.56 & &    0.09 & &    2448897.6872\\
$R$ &  &  &  &  &  &   \\
  \  & &   22.32 & &    0.09 & &    2447441.6340\\
  \  & &   22.55 & &    0.09 & &    2447447.6764\\
  \  & &   22.06 & &    0.07 & &    2448510.7125\\
  \  & &   22.01 & &    0.07 & &    2448511.6674\\
  \  & &   22.05 & &    0.06 & &    2448513.7535\\
  \  & &   21.91 & &   0.17 & &    2448977.5622\\
  \  & &   22.03 & &    0.04 & &    2448947.5392\\
  \  & &   22.37 & &    0.07 & &    2448897.6998\\
  \  & &   22.58 & &    0.08 & &    2448898.6209\\
  \  & &   22.34 & &    0.09 & &    2446729.6097\\
  \  & &   21.85 & &    0.06 & &    2448512.6653\\
  \  & &   22.00 & &    0.06 & &    2448509.8576\\
  \  & &   22.04 & &    0.07 & &    2446406.5993\\
  \  & &   22.09 & &    0.06 & &    2446408.5833\\
  \  & &   22.10 & &    0.08 & &    2447124.6479\\
  \  & &   21.99 & &    0.08 & &    2446744.6368\\
  \  & &   22.44 & &   0.12 & &    2446329.6562\\
  \  & &   21.93 & &    0.04 & &    2448946.7196\\
$I$ &  &  &  &  &  &    \\
  \  & &   21.99 & &   0.14 & &    2447441.6396\\
  \  & &   21.78 & &   0.12 & &    2447447.6819\\
  \  & &   21.47 & &    0.08 & &    2448510.7201\\
  \  & &   21.41 & &    0.08 & &    2448511.6583\\
  \  & &   21.53 & &    0.08 & &    2448513.6340\\
  \  & &   21.51 & &   0.17 & &    2448977.5759\\
  \  & &   21.61 & &    0.06 & &    2448947.5525\\
  \  & &   22.04 & &   0.14 & &    2448897.7084\\
  \  & &   21.78 & &   0.14 & &    2448898.6084\\
  \  & &   21.94 & &   0.21 & &    2446729.6139\\
  \  & &   21.55 & &   0.11 & &    2446407.6347\\
  \  & &   21.74 & &   0.18 & &    2447124.6438\\
  \  & &   21.49 & &   0.15 & &    2446744.6417\\
  \  & &   21.56 & &   0.12 & &    2447090.5882\\
  \  & &   21.63 & &   0.14 & &    2446330.6611\\
  \  & &   21.53 & &    0.07 & &    2448946.7350\\
\enddata
\end{deluxetable}

\begin{deluxetable}{cclclcl}
\tablecolumns{7}
\tablenum{9}
\tablewidth{0pt}
\tablecaption{Observations of C6 (P = 30.1 days) \label{tbl9}}
\tablehead{
\colhead{Filter} & \colhead{}  & \colhead{Mag.} & \colhead{} & \colhead{Error}
& \colhead{} &  \colhead{Julian Day}\\
}
\startdata
$B$  &  &  &  &  &  &    \\
  \  & &   22.77 & &    0.05 & &    2447441.6653\\
  \  & &   22.86 & &    0.10 & &  2448510.7882\\
  \  & &   23.06 & &    0.09 & &    2448511.6944\\
  \  & &   22.88 & &    0.10 & &  2447062.6361\\
  \  & &   22.35 & &    0.05 & &    2447472.6167\\
  \  & &   23.29 & &   0.18 & &    2446744.6611\\
  \  & &   22.77 & &    0.08 & &    2447090.6125\\
  \  & &   22.22 & &    0.09 & &    2447414.8458\\
  \  & &   22.45 & &    0.05 & &    2447419.7062\\
  \  & &   22.26 & &    0.08 & &    2447081.6049\\
  \  & &   22.56 & &    0.09 & &    2448976.6208\\
  \  & &   22.24 & &   0.12 & &    2448978.6345\\
  \  & &   22.13 & &    0.09 & &    2448947.6249\\
  \  & &   22.96 & &   0.17 & &    2448897.7201\\
$V$ &  &  &  &  &  &    \\
  \  & &   22.28 & &    0.04 & &    2447441.6590\\
  \  & &   22.25 & &   0.11 & &    2448510.7826\\
  \  & &   22.64 & &    0.10 & &  2448511.7069\\
  \  & &   22.53 & &    0.08 & &    2448512.7764\\
  \  & &   22.65 & &    0.10 & &  2448513.7483\\
  \  & &   22.23 & &    0.10 & &    2448509.8903\\
  \  & &   22.22 & &    0.06 & &    2447090.6083\\
  \  & &   21.88 & &    0.08 & &    2447081.6007\\
  \  & &   22.57 & &    0.09 & &    2447124.6708\\
  \  & &   22.03 & &    0.05 & &    2447472.6229\\
  \  & &   22.95 & &   0.19 & &    2446744.6569\\
  \  & &   22.36 & &    0.08 & &    2446729.6292\\
  \  & &   21.90 & &    0.09 & &    2447414.8403\\
  \  & &   22.04 & &    0.04 & &    2447419.7021\\
  \  & &   22.19 & &    0.08 & &    2447062.6458\\
  \  & &   21.97 & &    0.04 & &    2447447.7035\\
  \  & &   22.00 & &   0.11 & &    2448976.6353\\
  \  & &   21.93 & &    0.10 & &  2448978.6484\\
  \  & &   21.98 & &    0.04 & &    2448947.5844\\
  \  & &   22.32 & &    0.07 & &    2448897.7311\\
  \  & &   22.12 & &    0.06 & &    2448898.6789\\
$R$ &  &  &  &  &  &    \\
  \  & &   21.94 & &    0.04 & &    2447441.6528\\
  \  & &   21.82 & &    0.05 & &    2448510.7583\\
  \  & &   22.04 & &    0.08 & &    2448511.7153\\
  \  & &   21.91 & &    0.05 & &    2448512.7781\\
  \  & &   21.88 & &    0.05 & &    2448513.7521\\
  \  & &   21.72 & &    0.05 & &    2448509.8750\\
  \  & &   21.88 & &    0.06 & &    2447124.6764\\
  \  & &   22.39 & &   0.12 & &    2446744.6521\\
  \  & &   21.76 & &    0.05 & &    2447090.6000\\
  \  & &   21.68 & &    0.05 & &    2446729.6201\\
  \  & &   21.50 & &    0.08 & &    2447081.5917\\
  \  & &   21.60 & &    0.04 & &    2447447.6965\\
  \  & &   21.50 & &    0.09 & &    2448976.6483\\
  \  & &   21.60 & &    0.03 & &    2448947.5974\\
  \  & &   21.70 & &    0.05 & &    2448897.7402\\
  \  & &   21.74 & &    0.05 & &    2448898.6700\\
$I$ &  &  &  &  &  &    \\
  \  & &   21.54 & &    0.08 & &    2447441.6479\\
  \  & &   21.59 & &   0.11 & &    2448510.7417\\
  \  & &   21.38 & &    0.07 & &    2448511.7229\\
  \  & &   21.66 & &   0.12 & &    2448513.5979\\
  \  & &   21.55 & &    0.06 & &    2448512.7778\\
  \  & &   21.42 & &   0.16 & &    2447124.6812\\
  \  & &   21.55 & &   0.15 & &    2447090.5958\\
  \  & &   21.58 & &   0.16 & &    2446729.6243\\
  \  & &   21.32 & &    0.07 & &    2447447.6910\\
  \  & &   21.15 & &    0.09 & &    2448976.6609\\
  \  & &   21.32 & &    0.05 & &    2448947.6100\\
  \  & &   21.53 & &    0.09 & &    2448897.7494\\
  \  & &   21.44 & &    0.06 & &    2448898.6607\\
\enddata
\end{deluxetable}

\begin{deluxetable}{cclclcl}
\tablecolumns{7}
\tablenum{10}
\tablewidth{0pt}
\tablecaption{Observations of C7 (P = 26.1 days) \label{tbl10}}
\tablehead{
\colhead{Filter} & \colhead{}  & \colhead{Mag.} & \colhead{} & \colhead{Error}
& \colhead{} &  \colhead{Julian Day}\\
}
\startdata
$B$  &  &  &  &  &  &    \\
  \  & &   23.50 & &    0.07 & &    2447441.6878\\
  \  & &   23.00 & &    0.05 & &    2446407.6021\\
  \  & &   22.99 & &    0.07 & &    2447472.5924\\
  \  & &   23.27 & &    0.10 & &  2447090.5611\\
  \  & &   23.75 & &   0.11 & &    2446728.6868\\
  \  & &   22.81 & &    0.05 & &    2447419.6694\\
  \  & &   22.98 & &   0.11 & &    2446743.6097\\
  \  & &   23.32 & &   0.13 & &    2448512.7121\\
  \  & &   22.68 & &   0.13 & &    2448513.6222\\
$V$ &  &  &  &  &  &    \\
  \  & &   23.19 & &    0.10 & &  2447441.6847\\
  \  & &   22.40 & &    0.05 & &    2446407.6208\\
  \  & &   22.72 & &   0.12 & &    2447472.5979\\
  \  & &   22.88 & &    0.09 & &    2447090.5708\\
  \  & &   22.96 & &   0.12 & &    2446728.6819\\
  \  & &   22.81 & &   0.20 & &    2447414.7306\\
  \  & &   22.55 & &    0.07 & &    2447419.6653\\
  \  & &   22.95 & &   0.18 & &    2447062.5917\\
  \  & &   22.50 & &   0.13 & &    2446743.6229\\
  \  & &   22.68 & &   0.12 & &    2448512.7024\\
  \  & &   22.53 & &   0.13 & &    2448513.6168\\
  \  & &   23.00 & &   0.12 & &    2448946.6490\\
  \  & &   23.23 & &   0.17 & &    2448898.8294\\
  \  & &   22.45 & &    0.06 & &    2446406.6125\\
  \  & &   22.47 & &    0.06 & &    2446408.6236\\
  \  & &   23.19 & &   0.15 & &    2447124.6278\\
  \  & &   22.43 & &   0.13 & &    2446743.6250\\
  \  & &   23.06 & &   0.12 & &    2446729.5938\\
  \  & &   22.86 & &   0.18 & &    2446331.5681\\
  \  & &   22.40 & &    0.05 & &    2447447.6328\\
  \  & &   23.04 & &   0.18 & &    2448978.5743\\
  \  & &   23.26 & &   0.18 & &    2448509.8903\\
$R$ &  &  &  &  &  &   \\
  \  & &   22.66 & &    0.05 & &    2447441.6910\\
  \  & &   22.29 & &    0.09 & &    2447090.5514\\
  \  & &   22.25 & &   0.13 & &    2447062.5854\\
  \  & &   22.33 & &    0.09 & &    2448512.6941\\
  \  & &   22.31 & &    0.09 & &    2448513.6089\\
  \  & &   22.40 & &    0.07 & &    2448946.6623\\
  \  & &   22.85 & &    0.10 & &  2448897.6596\\
  \  & &   22.55 & &    0.06 & &    2446029.5344\\
  \  & &   22.14 & &    0.06 & &    2446406.6069\\
  \  & &   22.13 & &    0.06 & &    2446408.6146\\
  \  & &   22.46 & &    0.09 & &    2447124.6236\\
  \  & &   22.26 & &    0.10 & &  2446744.6194\\
  \  & &   22.53 & &    0.09 & &    2446729.5833\\
  \  & &   22.19 & &    0.08 & &    2446328.6771\\
  \  & &   22.57 & &    0.13 & &    2446329.6389\\
  \  & &   22.10 & &    0.04 & &    2447447.6472\\
  \  & &   22.66 & &   0.02 & &    2448978.6041\\
  \  & &   22.94 & &   0.13 & &    2448509.8810\\
$I$ &  &  &  &  &  &    \\
  \  & &   22.51 & &   0.15 & &    2447441.6965\\
  \  & &   22.14 & &   0.02 & &    2447090.5556\\
  \  & &   21.74 & &    0.09 & &    2448946.6748\\
  \  & &   22.24 & &   0.14 & &    2448897.6686\\
  \  & &   21.78 & &   0.13 & &    2446407.6104\\
  \  & &   21.68 & &   0.16 & &    2446744.6153\\
  \  & &   22.02 & &   0.19 & &    2446729.5875\\
  \  & &   21.83 & &   0.11 & &    2447419.6590\\
  \  & &   21.93 & &   0.13 & &    2447447.6528\\
\enddata
\end{deluxetable}

\begin{deluxetable}{cclclcl}
\tablecolumns{7}
\tablenum{11}
\tablewidth{0pt}
\tablecaption{Observations of C8 (P = 22.3 days) \label{tbl11}}
\tablehead{
\colhead{Filter} & \colhead{}  & \colhead{Mag.} & \colhead{} & \colhead{Error}
& \colhead{} &  \colhead{Julian Day}\\
}
\startdata
$B$  &  &  &  &  &  &    \\
  \  & &   23.76 & &    0.09 & &    2447441.6653\\
  \  & &   23.57 & &   0.11 & &    2448511.6944\\
  \  & &   23.97 & &   0.02 & &    2447062.6361\\
 \  & &   24.18 & &    0.02 & &    2447472.6167\\
  \  & &   22.63 & &    0.10 & &  2446744.6611\\
  \  & &   23.84 & &   0.18 & &    2447090.6125\\
  \  & &   23.65 & &   0.13 & &    2447419.7062\\
$V$ &  &  &  &  &  &    \\
  \  & &   22.82 & &    0.07 & &    2447441.6590\\
  \  & &   22.77 & &   0.15 & &    2448510.7826\\
  \  & &   23.10 & &   0.03 & &    2448511.7069\\
  \  & &   22.94 & &   0.11 & &    2448512.7764\\
  \  & &   23.07 & &   0.15 & &    2448513.7483\\
  \  & &   22.88 & &   0.18 & &    2448509.8903\\
  \  & &   23.29 & &   0.15 & &    2447090.6083\\
  \  & &   22.51 & &   0.14 & &    2447081.6007\\
  \  & &   22.25 & &    0.07 & &    2447124.6708\\
  \  & &   23.22 & &   0.15 & &    2447472.6229\\
  \  & &   22.28 & &    0.09 & &    2446744.6569\\
  \  & &   23.22 & &   0.15 & &    2446729.6292\\
  \  & &   22.92 & &    0.08 & &    2447419.7021\\
  \  & &   22.80 & &   0.12 & &    2447062.6458\\
  \  & &   23.25 & &   0.15 & &    2447447.7035\\
  \  & &   22.47 & &   0.13 & &    2448978.6484\\
  \  & &   23.26 & &   0.12 & &    2448947.5844\\
  \  & &   23.42 & &   0.18 & &    2448897.7311\\
  \  & &   23.34 & &   0.19 & &    2448898.6789\\
$R$ &  &  &  &  &  &    \\
  \  & &   22.37 & &    0.05 & &    2447441.6528\\
  \  & &   22.18 & &    0.10 & &  2448510.7583\\
  \  & &   22.51 & &   0.12 & &    2448511.7153\\
  \  & &   22.44 & &    0.09 & &    2448512.7781\\
  \  & &   22.54 & &    0.10 & &  2448513.7521\\
  \  & &   22.35 & &    0.09 & &    2448509.8750\\
  \  & &   21.88 & &    0.06 & &    2447124.6764\\
  \  & &   22.05 & &    0.08 & &    2446744.6521\\
  \  & &   22.81 & &    0.10 & &  2447090.6\\
  \  & &   22.53 & &    0.10 & &    2446729.6201\\
  \  & &   22.32 & &   0.16 & &    2447081.5917\\
  \  & &   22.62 & &    0.08 & &    2447447.6965\\
  \  & &   22.04 & &   0.14 & &    2448978.6613\\
  \  & &   22.94 & &   0.12 & &    2448947.5974\\
  \  & &   22.62 & &    0.10 & &  2448897.7402\\
$I$ &  &  &  &  &  &    \\
  \  & &   21.98 & &    0.10 & &  2447441.6479\\
  \  & &   22.05 & &   0.16 & &    2448510.7417\\
  \  & &   21.92 & &   0.13 & &    2448511.7229\\
  \  & &   21.90 & &    0.09 & &    2448512.7778\\
  \  & &   21.36 & &   0.14 & &    2447124.6812\\
  \  & &   22.13 & &   0.29 & &    2446744.6479\\
  \  & &   21.89 & &   0.18 & &    2446729.6243\\
  \  & &   21.59 & &   0.18 & &    2447414.8340\\
  \  & &   22.37 & &   0.16 & &    2447447.6910\\
  \  & &   22.35 & &   0.13 & &    2448947.6100\\
\enddata
\end{deluxetable}

\begin{deluxetable}{cclclcl}
\tablecolumns{7}
\tablenum{12}
\tablewidth{0pt}
\tablecaption{Observations of C9 (P= 20.2 days) \label{tbl12}}
\tablehead{
\colhead{Filter} & \colhead{}  & \colhead{Mag.} & \colhead{} & \colhead{Error}
& \colhead{} &  \colhead{Julian Day}\\
}
\startdata
$B$  &  &  &  &  &  &   \\
  \  & &   23.88 & &   0.14 & &    2447441.6878\\
  \  & &   23.74 & &   0.13 & &    2446407.6021\\
  \  & &   23.60 & &   0.17 & &    2447120.5347\\
  \  & &   23.80 & &   0.17 & &    2447472.5924\\
  \  & &   23.65 & &   0.17 & &    2447090.5611\\
  \  & &   23.90 & &   0.19 & &    2446728.6868\\
  \  & &   23.83 & &   0.17 & &    2447419.6694\\
  \  & &   22.82 & &   0.12 & &    2447062.6007\\
  \  & &   23.27 & &   0.16 & &    2446743.6097\\
  \  & &   23.04 & &   0.15 & &    2448946.6891\\
$V$ &  &  &  &  &  &    \\
  \  & &   23.25 & &   0.11 & &    2447441.6847\\
  \  & &   23.31 & &   0.14 & &    2446407.6208\\
  \  & &   23.27 & &   0.18 & &    2447120.5403\\
  \  & &   22.73 & &   0.12 & &    2447472.5979\\
  \  & &   23.06 & &   0.15 & &    2447090.5708\\
  \  & &   23.18 & &   0.13 & &    2446728.6819\\
  \  & &   22.84 & &   0.19 & &    2447414.7306\\
  \  & &   23.35 & &   0.16 & &    2447419.6653\\
  \  & &   22.49 & &   0.12 & &    2447062.5917\\
  \  & &   22.81 & &   0.17 & &    2446743.6229\\
  \  & &   22.49 & &    0.06 & &    2448946.6490\\
  \  & &   23.26 & &   0.16 & &    2446406.6125\\
  \  & &   23.40 & &   0.17 & &    2446408.6236\\
  \  & &   22.53 & &   0.11 & &    2447124.6278\\
  \  & &   22.69 & &   0.15 & &    2446743.6250\\
  \  & &   23.46 & &   0.20 & &    2446729.5938\\
  \  & &   22.60 & &    0.07 & &    2447447.6328\\
$R$ &  &  &  &  &  &    \\
  \  & &   22.86 & &    0.10 & &  2447441.6910\\
  \  & &   22.63 & &   0.18 & &    2447090.5514\\
  \  & &   22.62 & &   0.12 & &    2447414.7417\\
  \  & &   22.40 & &   0.15 & &    2447062.5854\\
  \  & &   22.43 & &    0.06 & &    2448946.6623\\
  \  & &   22.41 & &    0.07 & &    2446029.5344\\
  \  & &   22.87 & &   0.15 & &    2446406.6069\\
  \  & &   22.91 & &   0.14 & &    2446408.6146\\
  \  & &   22.36 & &    0.10 & &  2447124.6236\\
  \  & &   22.48 & &   0.15 & &    2446744.6194\\
  \  & &   22.72 & &   0.12 & &    2446729.5833\\
  \  & &   22.75 & &   0.16 & &    2446329.6389\\
  \  & &   22.30 & &    0.06 & &    2447447.6472\\
\enddata
\end{deluxetable}

\begin{deluxetable}{ccccccccccrcrcrcr}
\tablecolumns{17}
\tablenum{13}
\rotate\tablewidth{0pt}
\tablecaption{Properties of Unclassified Variables Found in NGC~0247\label{tbl13}}
\tablehead{
\colhead{ID} & \colhead{} & \colhead{Field} & \colhead{} & \multicolumn{3}{c}{Coordinates\tablenotemark{a}} 
& \colhead{} &
\colhead{P(?)} &  \colhead{} & \colhead{$<B>$} & \colhead{} & \colhead{$<V>$} & \colhead{} &
\colhead{$<R>$} & \colhead{} & \colhead{$<I>$} \\  
\colhead{} & \colhead{} & \colhead{} & \colhead{} & \colhead{x} & \colhead{} & \colhead{y} & \colhead{} & \colhead{(days)} &  \colhead{} 
& \colhead{$\sigma_{B}$} & \colhead{} & \colhead{$\sigma_{V}$} & \colhead{} & 
\colhead{$\sigma_{R}$} & \colhead{} & \colhead{$\sigma_{I}$} \\ 
}
\startdata
NGC~0247:[MF09]~P1 && 1 && 227.0 && 223.4 && 14.4 && (23.49) &&  21.56 &&  20.66 &&  19.59 \\
 &&  &&  &&  &&  && 0.03  &&  0.01  &&  0.02  &&  0.01 \\
NGC~0247:[MF09]~P2 && 3 && 100.6  && 305.6 && 16.3 &&  (22.59) &&  20.81  && 19.68  && 18.73 \\
   &&  &&  &&  &&  && 0.03 &&   0.01 &&   0.01  &&  0.01 \\
NGC~0247:[MF09]~P3 && 3 && 104.4 && 090.7 && 28.4 &&  (23.53) &&  21.49 &&  20.26 &&  19.14 \\
  &&  &&  &&  &&  &&  0.07  &&  0.03   && 0.03  &&  0.02 \\
NGC~0247:[MF09]~P4 && 3 && 052.4 && 311.8 && 30.8 &&  (22.84)  && 20.88 &&  19.89  && 18.93 \\
  &&  &&  &&  &&  &&  0.03  &&  0.02  &&  0.01 &&   0.01 \\
NGC~0247:[MF09]~P5 && 2 && 178.3 && 149.8 && 63.2 &&  (23.10) &&  22.09  && 21.54 &&  21.01 \\
  &&  &&  &&  &&  &&  0.05 &&   0.02 &&   0.02  &&  0.02 \\

\enddata
\tablenotetext{a}{Origin at bottom left (south-east) corner of frame}
\end{deluxetable}

\begin{deluxetable}{crcrrrrrcrrrrrrrr}
\tablecolumns{17}
\tablenum{14}
\rotate\tablewidth{8.0in}
\tablecaption{Properties of NGC~0247 Cepheids \label{tbl14}}
\tablehead{
\colhead{ID} & \colhead{} & \colhead{Field} & \colhead{} & \multicolumn{3}{c}{Coordinates\tablenotemark{a}\ $^{,}$\tablenotemark{b}} 
& \colhead{} &
\colhead{P} &  \colhead{} & \colhead{$<B>$} & \colhead{} & \colhead{$<V>$} & \colhead{} &
\colhead{$<R>$} & \colhead{} & \colhead{$<I>$} \\  
\colhead{} & \colhead{} & \colhead{} & \colhead{} & \colhead{x} & \colhead{} & \colhead{y} & \colhead{} & \colhead{(days)} &  \colhead{} 
& \colhead{$\sigma_{B}$} & \colhead{} & \colhead{$\sigma_{V}$} & \colhead{} & 
\colhead{$\sigma_{R}$} & \colhead{} & \colhead{$\sigma_{I}$} \\ 
}
\startdata
NGC~0247:[MF09]~C1 & & 1 & & 097.8  & & 474.5  & & 69.9 & & (22.31) & & 21.21 & & 20.65 & & 20.07\\
& & & & & & & & & &   0.08 & &  0.04  & & 0.03  & & 0.02\\
NGC~0247:[MF09]~C2 & & 1 & & 011.6  & & 250.5  & &  48.3 & & (22.77) & & 21.92 & & 21.51 & & 21.11\\
& & & & & & & & & & 0.13 & &  0.07 & &  0.05 & &  0.07\\
NGC~0247:[MF09]~C3 & & 2 & & 211.7  & & 302.1  & & 44.3 & & (22.99) & & 21.78 & & 21.81 & & 21.19\\
& & & & & & & & & &  0.13 & &  0.07 & &  0.05 & &  0.05\\
NGC~0247:[MF09]~C4 & & 2 & & 069.4 & &  290.9  & & 33.2 & & (22.86) & & 22.12 & & 21.79 & & 21.28\\
 & & & & & & & & & &  0.09 & &  0.04 & &  0.04 & &  0.02\\
NGC~0247:[MF09]~C5 & & 2 & & 151.1 & &  130.1  & & 31.0 & & (23.11) & & 22.59 & &   22.23 & & 21.68\\
& & & & & & & & & &   0.08 & &  0.05 & &  0.04 & &  0.04\\ 
NGC~0247:[MF09]~C6 & & 3 & & 055.5 & &  362.2  & &  30.1 & & (22.82) & & 22.37 & & 21.84 & & 21.44\\
 & & & & & & & & & &  0.11 & &  0.07 & &  0.06 & &  0.03\\
NGC~0247:[MF09]~C7 & & 1 & & 196.9 & &   201.3  & & 26.1 & & (23.30) & & 22.79 & & 22.36 & & 22.03\\
 & & & & & & & & & &  0.12 & &  0.06 & &  0.04 & &  0.09\\
NGC~0247:[MF09]~C8 & & 3 & & 044.7 & &  187.1  & & 22.3 & & (23.52) & & 22.93 & & 22.50 & & 22.06\\
 & & & & & & & & & &  0.20 & &  0.08 & &  0.08 & &  0.09\\
NGC~0247:[MF09]~C9 & & 1 & & 125.4 & &  208.0  & & 20.2 & & (23.49) & &  22.93 & & 22.57 & & \nodata \\  
& & & & & & & & & &   0.12 & &  0.09 & &  0.06 & & \nodata \\

\enddata
\tablenotetext{a}{Origin at bottom left (south-east) corner of frame}
\tablenotetext{b}{Scale = $0.6$ arcsec/pxl}
\end{deluxetable}

\begin{deluxetable}{crcrrrrrcrrrrrrrr}
\tablecolumns{17}
\tablenum{15}
\tablewidth{0pt}
\tablecaption{Revised VI Magnitudes and Updated Periods for Cepheids in NGC~0247 \label{tbl15}}
\tablehead{
\colhead{[MF09] ID} & \colhead{} & \colhead{[GV08] ID} & \colhead{} & \multicolumn{3}{}{} 
& \colhead{} &
\colhead{P} &  \colhead{} & \colhead{} & \colhead{} & \colhead{$<V>$} & \colhead{} &
\colhead{$<R>$} & \colhead{} & \colhead{$<I>$} \\  
\colhead{} & \colhead{} & \colhead{} & \colhead{} & \colhead{} & \colhead{} & \colhead{} & \colhead{} & \colhead{(days)} &  \colhead{} 
& \colhead{} & \colhead{} & \colhead{$\sigma_{V}$} & \colhead{} & 
\colhead{$\sigma_{R}$} & \colhead{} & \colhead{$\sigma_{I}$} \\ 
}
\startdata
NGC~0247:[MF09]~C1 & & cep020 & &  & & & & 69.9 & & & & 21.21 & & 20.65 & & 20.07\\
& & & & & & & & & & & &  0.03  & & 0.03  & & 0.02\\
NGC~0247:[MF09]~C2 & & cep017 & &  & & & &  48.3 & & & & 21.92 & & 21.51 & & 21.11\\
& & & & & & & & & & & &  0.07 & &  0.05 & &  0.07\\
NGC~0247:[MF09]~C3 & & cep016 & &  & & & & 44.38 & & & & 22.26 & & 21.83 & & 21.23\\
& & & & & & & & & & & &  0.06 & &  0.05 & &  0.05\\
NGC~0247:[MF09]~C4 & & cep011 & & & & & & 33.23 & & & & 22.12 & & 21.79 & & 21.28\\
 & & & & & & & & & & & &  0.04 & &  0.04 & &  0.02\\
NGC~0247:[MF09]~C5 & & cep008 & & & & & & 30.931 & & & & 22.59 & &   22.23 & & 21.68\\
& & & & & & & & & & & &  0.03 & &  0.05 & &  0.03\\ 
NGC~0247:[MF09]~C6 & & cep005 & & & & & &  27.785 & & & & 22.42 & & 21.77 & & 21.65\\
 & & & & & & & & & & & &  0.03 & &  0.06 & &  0.03\\
NGC~0247:[MF09]~C7 & &  & & & & & & 26.1 & & & & 22.76 & & 22.40 & & 21.96\\
 & & & & & & & & & & & &  0.06 & &  0.06 & &  0.09\\
NGC~0247:[MF09]~C8 & &  & & & &  & & 22.3 & & & & 22.88 & & 22.38 & & 21.91\\
 & & & & & & & & & & & &  0.08 & &  0.08 & &  0.10\\
NGC~0247:[MF09]~C9 & &  & & & &  & & 20.2 & & & &  22.93 & & 22.57 & & \nodata \\  
& & & & & & & & & & & &  0.09 & &  0.06 & & \nodata \\

\enddata
\end{deluxetable}

\begin{deluxetable}{lcccc}
\tablenum{16}
\tablewidth{0pc}        
\tablecaption{Summary of Cepheid Distance Moduli to NGC~0247\label{tbl16}}
\tablehead{
\colhead{Sample }      & \colhead{No. \& Bands}   &
\colhead{$\mu_o$ ($\sigma$)}          & \colhead{D(Mpc) ($\sigma$)}       &
\colhead{E(V-I) ($\sigma$)}}
\startdata
Original [MF09] VRI Sample     &       9 VRI  & 27.78 (0.13)        &    3.60 (0.22)   &   0.07 (0.05)      \\
Updated [MF09] VRI Sample      &       9 VRI  & 27.81 (0.06)        &    3.65 (0.17)   &   0.07 (0.04)      \\
Updated [MF09] VI Sample       &       9 VI   & 27.79 (0.13)        &    3.61 (0.23)   &   0.07 (0.04)      \\
     &       &        &      &         \\
Original [GV08] VI Sample      &      17 VI   & 27.82 (0.08)        &    3.66 (0.14)   &   0.15 (0.03)      \\
Updated [GV08] VI Sample       &      17 VI   & 27.87 (0.09)        &    3.75 (0.15)   &   0.11 (0.03)      \\
     &       &        &      &         \\
Updated \& VI Merged Sample    &      20 VI   & 27.85 (0.13)        &    3.72 (0.15)   &   0.11 (0.03)      \\
\enddata
\end{deluxetable}

\end{document}